\documentclass[aps,prb,twocolumn,amsmath,amssymb,
superscriptaddress,floatfix,nofootinbib,eqsecnum]{revtex4-2}
\usepackage{amsthm}
\usepackage{amsmath}
\usepackage{graphicx}
\usepackage{amssymb}
\usepackage{bm} 
\usepackage{flafter}
\usepackage{slashed} 
\usepackage{color}
\usepackage{bbm}
\usepackage{amsfonts}
\usepackage{cases}
\usepackage{float}
\usepackage{graphicx}
\usepackage{subfigure}
\usepackage{url}
\usepackage{framed}
\usepackage{cancel,soul,ulem}
\usepackage{appendix}
\usepackage{mathtools} 
\usepackage[breaklinks=true,colorlinks,citecolor=blue,linkcolor=blue,urlcolor=blue]{hyperref}

\def\=={\raisebox{0.35pt}{$\mathrm{:}$}\!\!=}

\newcommand{\bea}{\begin{eqnarray}}
\newcommand{\eea}{\end{eqnarray}}

\theoremstyle{definition}

\theoremstyle{remark}

\begin{document}

\title{
Deconfined classical criticality in the anisotropic quantum spin-{$\frac{1}{2}$}
XY model on the square lattice  
}

\author{Christopher Mudry}
\affiliation{
Condensed Matter Theory Group,
PSI Center for Scientific Computing,
Theory and Data,
5232 Villigen PSI, Switzerland
            }
\affiliation{
Institut de Physique, EPF Lausanne, CH-1015 Lausanne, Switzerland
            }  

\author{\"{O}mer M. Aksoy}
\affiliation{
Department of Physics, Massachusetts Institute of Technology, Cambridge,
{Massachusetts} 02139, USA
            }
  
\author{Claudio Chamon}
\affiliation{
Department of Physics, Boston University, Boston,
{Massachusetts} 02215, USA
            }

\author{Akira Furusaki}
\affiliation{
RIKEN Center for Emergent Matter Science, Wako, Saitama, 351-0198, Japan
            }
\affiliation{
Condensed Matter Theory Laboratory, RIKEN, Wako, Saitama, 351-0198, Japan
            }

\date{\today}

\begin{abstract}
The anisotropic quantum spin-{$\frac{1}{2}$} XY model
on a linear chain was solved by
Lieb, Schultz, and Mattis in 1961
{[Ann.\ Phys.\ \textbf{16}, 407 (1961)]}
and shown to display a continuous quantum phase transition
at the O(2) symmetric point
separating two gapped phases with competing Ising long-range order.
For the square lattice, the following is known.
The two competing Ising ordered phases
extend to finite temperatures,
up to a boundary where a transition to
the paramagnetic phase occurs,
and meet at the O(2) symmetric critical line along the
temperature axis
that ends at a tricritical point at the Berezinskii-Kosterlitz-Thouless
transition temperature where
the two competing phases meet the paramagnetic phase. 
We show that the first-order zero-temperature (quantum)
phase transition that separates the
competing phases as a function of the anisotropy parameter
is smoothed by thermal fluctuations into deconfined classical criticality.
\end{abstract}

\maketitle


\section{Introduction}
\label{sec:Introduction}

The anisotropic quantum spin-{$\frac{1}{2}$} XY model 
is defined on a lattice $\Lambda$ by the Hamiltonian
\begin{equation}
\hat{H}^{\,}_{\Lambda}(\gamma)\coloneqq
J\,\sum_{\langle jj'\rangle}
\left[
(1-\gamma)\,
\hat{S}^{x}_{j}\,
\hat{S}^{x}_{j'}
+
(1+\gamma)\,
\hat{S}^{y}_{j}\,
\hat{S}^{y}_{j'}
\right],
\label{eq:def anisotropic quantum spin-1/2 XY Lambda}
\end{equation}
where $\hat{S}^{x}_{j}$ and $\hat{S}^{y}_{j}$ are the {$x$} and {$y$}
components of the quantum spin-{$\frac{1}{2}$} operator at site $j$, and
energies are measured in units of $J>0$.  The notation
$\langle jj'\rangle$ refers to a directed pair of nearest-neighbor
sites $j,j'\in\Lambda$.  The anisotropy is quantified by the parameter
$-1\leq\gamma\leq+1$.

\begin{figure}[t!]
(a)
\includegraphics[width=0.8\columnwidth]{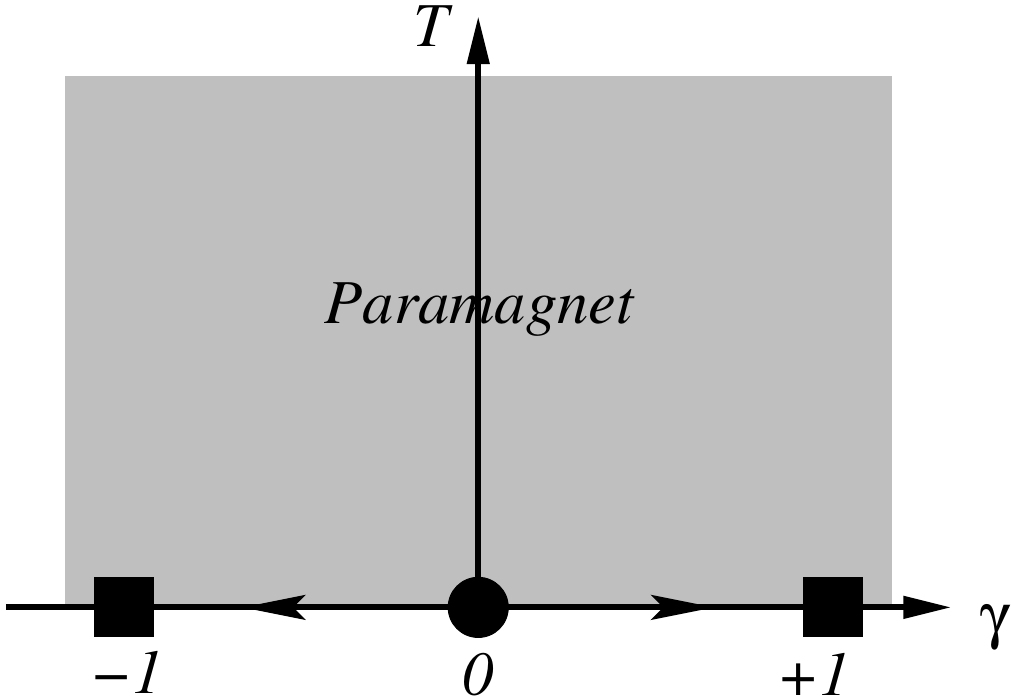}
\\\medskip
(b)  
\includegraphics[width=0.8\columnwidth]{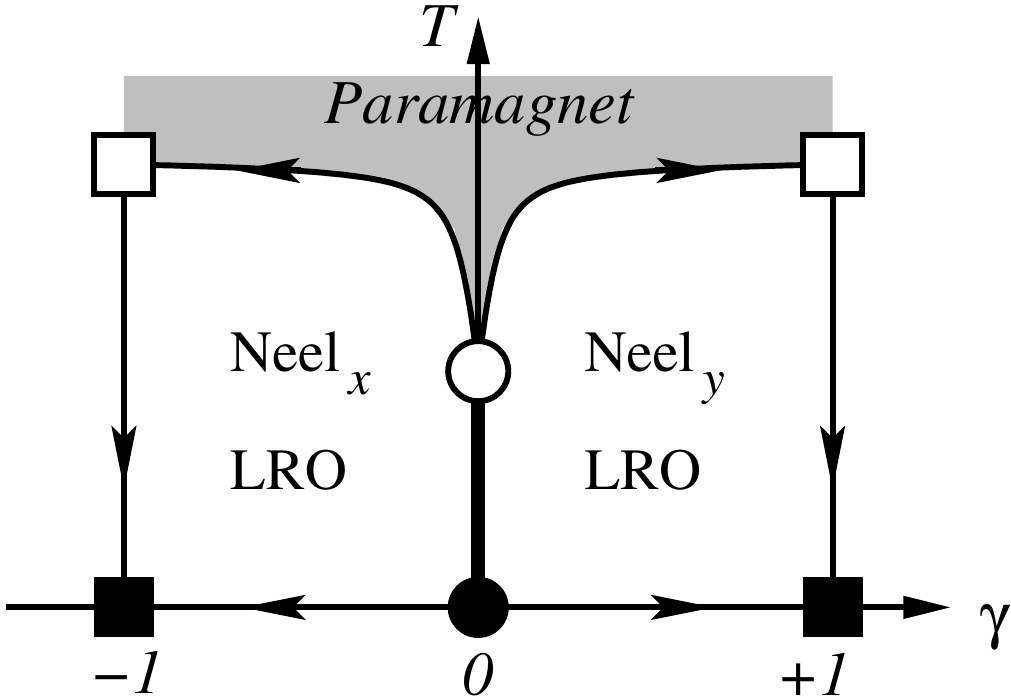}
\caption{
(a) Phase diagram of the one-dimensional anisotropic
quantum spin-{$\frac{1}{2}$} XY model defined by Hamiltonian    
(\ref{eq:def anisotropic quantum spin-1/2 XY Lambda})
with $\Lambda$ a linear chain.
The black squares represent the two attractive fixed points
for which the twofold-degenerate ground states are the classical N\'eel states
with the staggered magnetization per site either parallel or antiparallel
to the $x$ ($y$) direction for $\gamma=-1$ ($\gamma=+1$)
in spin space. The black circle represents the
quantum critical point realizing a {$\mathsf{c}=1$}
CFT with $\mathrm{U}(1)$ current algebra.
(b) Phase diagram of the two-dimensional anisotropic
quantum spin-{$\frac{1}{2}$} XY model defined by Hamiltonian    
(\ref{eq:def anisotropic quantum spin-1/2 XY Lambda})
with $\Lambda$ a square lattice.
The black squares have the same interpretation as in
(a). The white squares are the Ising transition temperature
for the classical antiferromagnetic Ising model on the square lattice.
The tricritical point at which the 
N\'eel${}^{\,}_{x}$,
N\'eel${}^{\,}_{y}$,
and paramagnetic phases meet is the
BKT transition temperature.
It is connected to the black circle
by a line of critical points, each one of which
realizes a $\mathrm{O}(2)$ symmetric
algebraic phase of matter.
The black circle realizes a first-order
phase boundary at zero temperature
as it is characterized by a non-vanishing
staggered magnetization per site that
can point in any direction along
the unit circle in spin space.
The arrows represent the renormalization group flows
in both (a) and (b).
}
\label{fig:phase diagrams d=1 and d=2}
\end{figure}

When the lattice $\Lambda$ is a linear chain,
Lieb, Schultz, and Mattis \cite{Lieb61}
expressed the spectrum of Hamiltonian
(\ref{eq:def anisotropic quantum spin-1/2 XY Lambda})
in closed analytical form 
by mapping $\hat{H}^{\,}_{\Lambda}(\gamma)$
through the Jordan-Wigner transformation \cite{Jordan28}
to a non-interacting nearest-neighbor one-dimensional tight-binding Hamiltonian
written in terms of Majorana fermions.
They showed that, in the thermodynamic limit,
the ground state is twofold degenerate and gapped
for any non-vanishing anisotropy
$\gamma\neq0$ with a non-vanishing expectation value for
the staggered magnetization per site pointing along the $x$ ($y$) direction
in spin space for $-1\leq\gamma<0$ ($0<\gamma\leq+1$)
and whose amplitude is maximal at $\gamma=\pm1$,
while it vanishes in a monotonic and continuous way upon approaching
$\gamma=0$. At the $\mathrm{O}(2)$ symmetric point $\gamma=0$
and after the thermodynamic limit has been taken,
Hamiltonian
(\ref{eq:def anisotropic quantum spin-1/2 XY Lambda})
realizes a quantum critical point 
that belongs to the universality class of a
{$\mathsf{c}=1$} conformal field theory (CFT) 
with $\mathrm{O}(2)$ current algebra.
The modern interpretation of this quantum critical point is (i)
that of a deconfined quantum critical point (DQCP) in the quantum spin-{$\frac{1}{2}$}
representation \cite{Mudry19}
(see also Refs.\ \cite{Jiang19} and \cite{Prakash24})
or (ii) that of a quantum topological critical point separating two
distinct fermionic invertible topological phases of matter in the Majorana
representation obtained from the Jordan-Wigner transformation
\cite{Verresen17,Aksoy24}. Thermal fluctuations destroy
the long-range ordered phases at any non-vanishing temperature
\cite{Ising25}.
The corresponding phase diagram can be found in Fig.\
\ref{fig:phase diagrams d=1 and d=2}(a).

The goal of this paper is to revisit the phase diagram of the
two-dimensional anisotropic quantum spin-{$\frac{1}{2}$} XY model, and to analyze
the model through the lens of deconfined criticality, i.e., continuous
quantum phase transitions between two gapped phases with competing
long-range order. We focus on Hamiltonian
(\ref{eq:def anisotropic quantum spin-1/2 XY Lambda})
with $\Lambda$ a square lattice. Our
main result is that each point along the phase boundary
$0<T\leq T^{\,}_{\mathrm{BKT}}$ when $\gamma=0$ realizes a deconfined
classical critical point in the universality class of a
{$\mathsf{c}=1$} CFT. This boundary terminates in a first-order
phase transition at zero temperature and in a
Berezinskii–Kosterlitz–Thouless (BKT) transition
\cite{Berezinsky71,Kosterlitz73} at the tricritical point
$T^{\,}_{\mathrm{BKT}}$. The corresponding phase diagram can be found
in Fig.\ \ref{fig:phase diagrams d=1 and d=2}(b).

A concise review of the literature dedicated to the
two-dimensional anisotropic quantum spin-{$\frac{1}{2}$} XY model is given in Sec.\
\ref{sec:A brief history when Lambda is a square lattice}.
The staggered magnetization per site is defined in Sec.\ 
\ref{sec:Order parameters}.
The first-order quantum phase transition between the competing
N\'eel${}^{\,}_{x}$
and
N\'eel${}^{\,}_{y}$
phases is derived in Sec.\
\ref{sec:Discontinuous quantum phase transition}.
The presence of deconfined classical criticality
at non-vanishing temperature is shown in
Sec.~\ref{sec:Deconfined classical criticality when ...}.
We close with an extension of the
two-dimensional anisotropic quantum spin-{$\frac{1}{2}$} XY model
to the two-dimensional anisotropic antiferromagnetic
quantum spin-{$\frac{1}{2}$} XYZ model in Sec.\
\ref{sec:Discussion}.

\section{A brief history when $\Lambda$ is a square lattice}
\label{sec:A brief history when Lambda is a square lattice}

The family (\ref{eq:def anisotropic quantum spin-1/2 XY Lambda})
of Hamiltonian labeled by the anisotropy parameter
$-1\leq\gamma\leq+1$
contains the following three Hamiltonians
\begin{subequations}
\label{eq:three limiting cases}
\begin{align}
&
\hat{H}^{\,}_{\mathrm{Ising}^{\,}_{x};\Lambda}\coloneqq
2J\;\sum_{\langle jj'\rangle}
\hat{S}^{x}_{j}\,
\hat{S}^{x}_{j'},
\label{eq:three limiting cases a}
\\
&
\hat{H}^{\,}_{\mathrm{Ising}^{\,}_{y};\Lambda}\coloneqq
2J\;\sum_{\langle jj'\rangle}
\hat{S}^{y}_{j}\,
\hat{S}^{y}_{j'},
\label{eq:three limiting cases b}
\\
&
\hat{H}^{\,}_{\mathrm{O}(2);\Lambda}\coloneqq
J\;\sum_{\langle jj'\rangle}
\left(
\hat{S}^{x}_{j}\,
\hat{S}^{x}_{j'}
+
\hat{S}^{y}_{j}\,
\hat{S}^{y}_{j'}
\right),
\label{eq:three limiting cases c}
\end{align}
\end{subequations}
when $\gamma=-1$, $\gamma=+1$, and $\gamma=0$, respectively.

The Ising-limit Hamiltonians (\ref{eq:three limiting cases a}) and
(\ref{eq:three limiting cases b}) are exactly solvable on a square
lattice $\Lambda$~\cite{Peierls36,Onsager44}; they display long-range
antiferromagnetic order along different axis, either the $x$ or
$y$ direction, respectively, in spin space.
The quantum spin-{$\frac{1}{2}$} XY Hamiltonian
(\ref{eq:three limiting cases c}) has the continuous U(1) symmetry of
the spin rotation about the $z$ axis and
was introduced by Matsubara and Matsuda in 1956 \cite{Matsubara56}.
The two-dimensional quantum spin-{$\frac{1}{2}$} XY Hamiltonian
(\ref{eq:three limiting cases c})
is not integrable when the lattice $\Lambda$ is the square lattice.
The phase diagram along the temperature axis of
the two-dimensional quantum spin-{$\frac{1}{2}$} XY Hamiltonian
(\ref{eq:three limiting cases c})
for a square lattice has been studied by combining
exact diagonalization, uncontrolled approximations,
partial rigorous results, and  Monte Carlo techniques
in Refs.\
\cite{Suzuki77,Pearson77,Betts77,Rogiers78a,Rogiers78b,Oitmaa78a,Oitmaa78b,Suzuki78,Uchinami79,Oitmaa80,Deraedt84a,Deraedt84b,Loh85,Takahashi86,Gomez-Santos87,Tang88,Okabe88,Kennedy88,Chakravarty89,Drzewinski89,Ding90a,Hamer91,Zhang92,Ding92a,Makivic92,Hasenfratz93,Ji94,Pires96,Harada98,Ying98,Hamer99,Sandvik99,Schindelin01,Melko04,Carrasquilla12,Hofmann14,Hofmann16,Bishop17},
following the discovery of the (classical)
Berezinskii–Kosterlitz–Thouless (BKT) transition
\cite{Berezinsky71,Kosterlitz73}.
One important goal of these studies was to understand the interplay between
quantum and thermal fluctuations and the fate of the BKT transition present
in the classical XY model on the square lattice.

The first question that could be answered rigorously was the fate
of the classical antiferromagnetic (i.e., the N\'eel)
ground state of the classical XY model on the square lattice $\Lambda$
with the classical Hamiltonian
\begin{equation}
\begin{split}
&
H^{\,}_{\mathrm{O}(2);\Lambda}\coloneqq
J\;\sum_{\langle jj'\rangle}
\left(
s^{x}_{j}\,
s^{x}_{j'}
+
s^{y}_{j}\,
s^{y}_{j'}
\right),
\\
&
\left(s^{x}_{j}\right)^{2}
+
\left(s^{y}_{j}\right)^{2}=1,
\qquad
s^{x}_{j},s^{y}_{j}\in\mathbb{R},
\end{split}
\label{eq:def classical XY model lattice Lambda}
\end{equation}
after trading $H^{\,}_{\mathrm{O}(2);\Lambda}$ for
the quantum spin-{$\frac{1}{2}$} XY Hamiltonian
(\ref{eq:three limiting cases c}).
It was shown by Kennedy, Lieb, and Shastry \cite{Kennedy88}
that the expectation value of the
staggered magnetization per site
in the ground state of
the quantum spin-{$\frac{1}{2}$} XY Hamiltonian
(\ref{eq:three limiting cases c})
is non-vanishing in the thermodynamic limit
and that it saturates to its classical N\'eel limit
if the quantum spin-{$\frac{1}{2}$} degrees of freedom are traded
for quantum spin $S$ with $S\in\mathbb{N}/2$
degrees of freedom in the limit $S\to\infty$.
By Goldstone theorem \cite{Nambu60,Goldstone61}, it then follows that
the spectrum of the quantum spin-{$\frac{1}{2}$} XY Hamiltonian
(\ref{eq:three limiting cases c})
must be gapless in the thermodynamic limit
owing to the presence of magnons (quantum spin waves)
at arbitrarily small energies
above that of the ground state.

An effective quantum field theory for
these magnons is that of the $\mathrm{O}(2)$ quantum non-linear sigma model
[$\mathrm{O}(2)$-QNLSM] in two-dimensional space
\cite{Chakravarty89,Hasenfratz93}.
This effective theory predicts that the thermal magnons downgrade
the long-range order to algebraic quasi-long-range order for
any non-vanishing yet not too large temperature.
Exact diagonalization studies \cite{Hamer99}
agree quantitatively with the predictions made from the
$\mathrm{O}(2)$-QNLSM for the
finite-size scaling behavior of the ground-state energy per site,
the spontaneous staggered magnetization per site,
the spin-wave velocity, and the spin-wave stiffness
at vanishing temperature.
The same is true for the low-temperature dependence
of observables such as magnetization, spin susceptibilities,
or spin-spin correlation functions obtained from
Monte-Carlo simulations~\cite{Sandvik99}.
All Monte-Carlo simulations after 1990
were also able to verify that the
algebraic phase turns 
into a paramagnetic phase
through a BKT transition
at the non-vanishing temperature
$T^{\,}_{\mathrm{BKT}}>0$.

None of the studies of the
two-dimensional anisotropic quantum spin-{$\frac{1}{2}$} XY Hamiltonians
(\ref{eq:def anisotropic quantum spin-1/2 XY Lambda})
when the lattice $\Lambda$ is the square lattice
(see Refs.\ \cite{Hamer91,Usman15,Wu16,Farajollahpour18,Zhao21})
address the following question.
What is the nature of the phase transition
when tuning $-1\leq\gamma\leq+1$
at fixed temperature $0\leq T<T^{\,}_{\mathrm{BKT}}$
through the phase boundary at $\gamma=0$?
This transition is not of the Landau-Ginzburg type
as the long-range ordered phases
on either side of the phase boundary,
the  
N\'eel${}^{\,}_{x}$
and
N\'eel${}^{\,}_{y}$
phases, break spontaneously distinct symmetries.
Is it a continuous phase transition or a discontinuous one?
We will demonstrate that the zero-temperature (quantum)
phase transition at $\gamma=0$ is discontinuous,
whereas at any nonzero temperature
$0<T<T^{\,}_{\mathrm{BKT}}$ the transition at $\gamma=0$ (i) is
continuous and (ii) serves as an example of a deconfined classical
criticality.

\section{Order parameters}
\label{sec:Order parameters}

Let $|\Lambda|$ denote the number of sites of the
square lattice $\Lambda$, whereby we shall take advantage of the fact
that $\Lambda$ can be partitioned into two square 
sublattices $\Lambda^{\,}_{A}$ and $\Lambda^{\,}_{B}$ that are
equal in cardinality,
disjoint, and such that the nearest neighbors to any
site from $\Lambda^{\,}_{A}$ belong to the sublattice $\Lambda^{\,}_{B}$
and vice versa.
Let $|j|$ assign the value $0$ ($1$) to $j$ if $j\in\Lambda$ belongs
to sublattice $\Lambda^{\,}_{A}$ ($\Lambda^{\,}_{B}$).
Let $\bm{B}\in\mathbb{R}^{2}$ denote a real-valued vector
$\bm{B}=(B^{x}, B^{y})$,
and let $\hat{\bm{S}}^{\,}_{j}$ denote the operator-valued
vector $\hat{\bm{S}}^{\,}_{j}=(\hat{S}^{x}_{j}, \hat{S}^{y}_{j})$
for any $j\in\Lambda$.
Define the partition function
\begin{equation}
Z^{\,}_{\Lambda}(\gamma,\beta;\bm{B})\coloneqq
\mathrm{Tr}
\left[  
e^{
-\beta\,
\hat{H}^{\,}_{\Lambda}(\gamma)
+\beta\,
\sum\limits_{j\in\Lambda}
(-1)^{|j|}\,
\bm{B}\cdot\hat{\bm{S}^{\,}_{j}}
}
\right]
\label{eq:def quantum partition function}
\end{equation}
at the inverse temperature $\beta\geq0$ (Boltzmann constant is set to unity).
The order parameter of the N\'eel$_{x,y}$ phase,
the staggered magnetization per site,
is defined by
\begin{equation}
\bm{m}^{\,}_{\mathrm{sta}}(\gamma,\beta)\coloneqq
\lim_{|\bm{B}|\to0}
\left\{
\lim_{|\Lambda|\to\infty}
\left[
\frac{1}{|\Lambda|}\,
\frac{1}{\beta}\,
\frac{\partial}{\partial\bm{B}}
\ln\,Z(\gamma,\bm{B})
\right]
\right\},
\label{eq:def staggered magnetization}
\end{equation}
where the order of limits matters.
To be more precise, in the spontaneously long-range ordered phases,
the sign of the projection of
$\bm{m}^{\,}_\mathrm{sta}(\gamma,\beta)$
along the $x$ or $y$ axis
depends on how the limit
$|\bm{B}|\to0$ is taken, i.e.,
with positive
$B^{x,y}\downarrow0$
or with negative
$B^{x,y}\uparrow0$.

For any given $-1\leq\gamma\leq+1$,
the magnitude
$|\bm{m}^{\,}_{\mathrm{sta}}(\gamma,\beta)|$
is maximal at $\beta=\infty$ (zero temperature)
and decreases with decreasing $\beta$.

For any given temperature $1/\beta$,
the magnitude
$|\bm{m}^{\,}_{\mathrm{sta}}(\gamma,\beta)|$
must be a decreasing function of $\gamma$
on the interval $-1\leq\gamma\leq0$
with its maximum at $\gamma=-1$ and
minimum at $\gamma=0$,
as quantum fluctuations increase with increasing $\gamma$.
Conversely, for any given temperature $1/\beta$,
the magnitude
$|\bm{m}^{\,}_{\mathrm{sta}}(\gamma,\beta)|$
must be an increasing function of $\gamma$
on the interval $0\leq\gamma\leq+1$
with its minimum at $\gamma=0$ and
maximum at $\gamma=+1$,
as quantum fluctuations decrease with increasing $\gamma$.
Quantum fluctuations are maximal at the fixed point
$\gamma=0$.

\section{Discontinuous quantum phase transition at zero temperature}
\label{sec:Discontinuous quantum phase transition}

At zero temperature, i.e., $\beta=\infty$,
the magnitude
$|\bm{m}^{\,}_{\mathrm{sta}}(\gamma,\infty)|$
is maximal at $\gamma=-1$ and $\gamma=+1$,
where it saturates to the value {$\frac{1}{2}$}, 
for
$\hat{H}^{\,}_{\Lambda}(\gamma)$
defined by Eq.\
(\ref{eq:def anisotropic quantum spin-1/2 XY Lambda})
simplifies to the classical Ising model
(\ref{eq:three limiting cases a}) at $\gamma=-1$
and
(\ref{eq:three limiting cases b}) at $\gamma=+1$, respectively.
More precisely,
\begin{equation}
\bm{m}^{\,}_{\mathrm{sta}}(-1,\infty)=
S\begin{pmatrix}\pm1\\0\end{pmatrix},
\quad
\bm{m}^{\,}_{\mathrm{sta}}(+1,\infty)=
S\begin{pmatrix}0\\\pm1\end{pmatrix},
\end{equation}
with $S=\frac{1}{2}$.
The $\pm$ sign represents twofold degenerate ground states
and corresponds to the two different ways alluded after
Eq.\ (\ref{eq:def staggered magnetization})  
for taking the limit $\bm{B}\to0$.

Let us first consider the interval $-1\leq\gamma\leq0$.
As long as the gap does not close,
the direction of the staggered magnetization is expected to
remain parallel to that of
$\bm{m}^{\,}_{\mathrm{sta}}(-1,\infty)$.
The spin-wave approximation from
Ref.\ \cite{Hamer91},
chiral perturbation theory from Ref.\ \cite{Hasenfratz93},
exact diagonalization,
and Monte-Carlo simulations
all find that the spectral gap between the twofold degenerate ground states
and all excited states is a monotonic decreasing function of $\gamma$
that vanishes at $\gamma=0$.
The exact result from
Kennedy, Lieb, and Shastry \cite{Kennedy88}
implies that the magnitude of
$|\bm{m}^{\,}_{\mathrm{sta}}(0,\infty)|$
is non-vanishing. Hence, we conclude that
$\bm{m}^{\,}_{\mathrm{sta}}(\gamma,\infty)$
is parallel to
$\bm{m}^{\,}_{\mathrm{sta}}(0,\infty)$
for any $-1\leq\gamma<0$
and converges in magnitude to
$|\bm{m}^{\,}_{\mathrm{sta}}(0,\infty)|$
in the limit $\gamma\uparrow0$.

The same analysis on the interval
$0\leq\gamma\leq+1$
predicts that
$\bm{m}^{\,}_{\mathrm{sta}}(\gamma,\infty)$
is parallel to
$\bm{m}^{\,}_{\mathrm{sta}}(+1,\infty)$
for any $0<\gamma\leq+1$
and converges 
in magnitude to
$|\bm{m}^{\,}_{\mathrm{sta}}(0,\infty)|$
in the limit $\gamma\downarrow0$.
At $\gamma=0$, any direction of
$\bm{m}^{\,}_{\mathrm{sta}}(0,\infty)$
is allowed
in the $S^{x}$-$S^{y}$ plane.

Whereas the magnitude
$|\bm{m}^{\,}_{\mathrm{sta}}(\gamma,\infty)|$
is continuous across the phase boundary
at $\gamma=0$,
$\bm{m}^{\,}_{\mathrm{sta}}(\gamma,\infty)$
is a discontinuous function at $\gamma=0$
as the direction of
$\bm{m}^{\,}_{\mathrm{sta}}(\gamma,\infty)$
changes in a discontinuous way across
the phase boundary at $\gamma=0$.
This discontinuity reflects a level crossing,
that is made possible by the closing of the
gap between the branch of magnons
(the pseudo Goldstone bosons in the terminology of Ref.\
\cite{Hasenfratz93})
and the two-fold degenerate ground states,
upon approaching the phase boundary at $\gamma=0$
from either the N\'eel${}_{x}$ or N\'eel${}_{y}$ side.

We have thus shown that the quantum phase boundary at
$\gamma=0$
realizes a discontinuous quantum phase transition
as a function of $\gamma$,
in spite of the magnon gap closing at the transition
and contrary to some claims 
\cite{Usman15,Wu16,Farajollahpour18,Zhao21}
made in the literature that the transition is continuous.

\section{Deconfined classical criticality when $\gamma=0$ at non-vanishing
temperature}
\label{sec:Deconfined classical criticality when ...}

\subsection{Staggered magnetization at non-vanishing temperature}

We shall take the two following empirical facts
(based on Monte-Carlo simulations) as given.
First, the N\'eel${}_{x}$ or N\'eel${}_{y}$ phases meet at the phase boundary
$\gamma=0$ for any given non-vanishing temperature lower than the
BKT critical temperature
$T^{\,}_{\mathrm{BKT}}=1/\beta^{\,}_{\mathrm{BKT}}$,
which itself is lower than
the critical temperature
$T^{\mathrm{2D}}_{\mathrm{Ising}}=1/\beta^{\mathrm{2D}}_{\mathrm{Ising}}$
of the classical Ising model on the square lattice.
Second, at any given finite value of 
$\beta>\beta^{\,}_{\mathrm{BKT}}$,
the strict monotonic dependence of
$|\bm{m}^{\,}_{\mathrm{sta}}(+\gamma,\beta)|=
|\bm{m}^{\,}_{\mathrm{sta}}(-\gamma,\beta)|$
on $-1\leq\gamma\leq0$ is qualitatively the same as that when $\beta=\infty$
except for one crucial difference, namely
\begin{equation}
|\bm{m}^{\,}_{\mathrm{sta}}(0,\beta)|=0.
\end{equation}
This difference opens the door for the possibility that
the phase boundary at $\gamma=0$ and
$\beta^{\,}_{\mathrm{BKT}}<\beta<\infty$
is a continuous phase transition.
We will discuss the power-law dependence of
$|\bm{m}^{\,}_{\mathrm{sta}}(\gamma,\beta)|$ for small $|\gamma|$,
after discussing the reduction of the QNLSM to the classical NLSM
at finite temperatures.

\subsection{Quantum Non-linear Sigma Model}

As shown by Chakravarty, Halperin, and Nelson \cite{Chakravarty89},
the effective field theory that describes the $\mathrm{O}(2)$ symmetric
line $\gamma=0$ for the regime of temperature
\begin{equation}
0\leq 1/\beta\lesssim J
\label{eq:condition low T}
\end{equation}
is the 
$\mathrm{O}(2)$ QNLSM defined by the $d\downarrow2$ and $N\downarrow2$
limits of the partition function
of the $d$-dimensional $\mathrm{O}(N)$ QNLSM{:}
\begin{widetext}
\begin{subequations}
\label{eq:d-dimensional O(N) QNLSM non zero T}
\begin{align}
&
Z^{\mathrm{eff}}_{\mathrm{O}(N)}
(\beta,\rho^{\,}_{\mathfrak{a}},c^{\,}_{\mathfrak{a}})\coloneqq
\left[
\prod_{0\leq\tau\leq\beta\,\hbar}\,
\prod_{\bm{r}\in\mathbb{R}^{d}}^{|\bm{r}|\geq\mathfrak{a}}\,
\int
\mathrm{d}^{N}\bm{n}(\tau,\bm{r})\,
\delta\bm{\big(}\bm{n}^{2}(\tau,\bm{r})-1\bm{\big)}
\right]
\exp
\left(
-
\frac{
	S^{\mathrm{eff}}_{\mathrm{O}(N)}(\beta,\rho^{\,}_{\mathfrak{a}},c^{\,}_{\mathfrak{a}})
}
{
	\hbar
}
\right),
\label{eq:d-dimensional O(N) QNLSM non zero T a}
\\
&
S^{\mathrm{eff}}_{\mathrm{O}(N)}
(\beta,\rho^{\,}_{\mathfrak{a}},c^{\,}_{\mathfrak{a}})\coloneqq
\frac{\rho^{\,}_{\mathfrak{a}}}{2}
\int\limits_{0}^{\beta\,\hbar}\mathrm{d}\tau
\int\limits_{|\bm{r}|\geq\mathfrak{a}}\mathrm{d}^{d}\bm{r}
\left[
\sum_{a=1}^{d}
\left(
\frac{\partial\bm{n}}{\partial r^{\,}_{a}}
\right)^{2}\!\!(\tau,\bm{r})
+
\frac{1}{c^{2}_{\mathfrak{a}}}
\left(
\frac{\partial\bm{n}}{\partial\tau}
\right)^{2}\!\!(\tau,\bm{r})
\right],
\label{eq:d-dimensional O(N) QNLSM non zero T b}
\end{align}
\end{subequations}
\end{widetext}
where the $N$-dimensional unit vector $\bm{n}(\tau,\bm{r})$
represents the soft fluctuations of the order parameter, i.e.,
the staggered magnetization
$\bm{m}^{\,}_{\mathrm{sta}}$.
The condition (\ref{eq:condition low T}) translates to
\begin{equation}
1\lesssim\beta\,\rho^{\,}_{\mathfrak{a}},
\label{eq:d-dimensional O(N) QNLSM non zero T c}
\end{equation}
and the periodic boundary conditions in imaginary time~$\tau$
\begin{equation}
\bm{n}(\tau+\beta\,\hbar,\bm{r})=
\bm{n}(\tau,\bm{r})
\label{eq:d-dimensional O(N) QNLSM non zero T d}
\end{equation}
must hold for all $\tau$ and $\bm{r}$.
We have made explicit the dependence of the effective partition function
on two bare couplings.
One is the spin stiffness $\rho^{\,}_{\mathfrak{a}}$
that carries the unit of energy times $(\mathrm{length})^{2-d}$;
its bare value being given by
$J\,S^{2}\,\mathfrak{a}^{2-d}$
with $S={\frac{1}{2}}$.
The other is the spin-wave velocity $c^{\,}_{\mathfrak{a}}$
that carries the unit of speed; its bare value being given by
$2\sqrt{d}\,J\,S\,\mathfrak{a}/\hbar$.
These couplings are assigned the subindex $\mathfrak{a}$
as their (non-universal) numerical values are set by the (non-unique)
underlying lattice regularization with characteristic lattice spacing
$\mathfrak{a}$ whose long wave length and low energy limit is
captured by the $d$-dimensional $\mathrm{O}(N)$-QNLSM.
The $\mathrm{O}(N)$ QNLSM (\ref{eq:d-dimensional O(N) QNLSM non zero T})
is a strongly interacting theory
owing to the constraint $\bm{n}^{2}=1$. The spin-wave approximation
consists in solving this constraint by writing
\begin{equation}
\bm{n}=\begin{pmatrix}\bm{\pi}\\\sigma\end{pmatrix},
\qquad
\sigma=+\sqrt{1-\bm{\pi}^{2}}>0,
\label{eq:d-dimensional O(N) QNLSM non zero T e}
\end{equation}
and demanding that the vector field (whose components are the spin waves)
$\bm{\pi}\in\mathbb{R}^{N-1}$
is a smooth vector field. This parametrization breaks the 
global $\mathrm{O}(N)$ symmetry explicitly down to $\mathrm{O}(N-1)$
since only the positive root is chosen for the $\sigma$ component.

\subsection{The phase boundary at $\gamma=0$ and $\beta=\infty$}

At zero temperature, i.e., $\beta=\infty$,
the effective theory (\ref{eq:d-dimensional O(N) QNLSM non zero T})
with $d=2$ and $N=2$
is in the same universality class
as the three-dimensional classical XY model defined by
the partition function with the Hamiltonian
(\ref{eq:def classical XY model lattice Lambda}),
whereby $\Lambda$ is a cubic lattice, at an effective
inverse temperature $\beta^{\,}_{\mathrm{eff}}$ chosen such that
the connected two-point spin-spin correlation functions decay
with the same correlation length $\xi$ at long separations
in both theories.
The interpretation of the correlation length $\xi$ is the following.
From the point of view of the
two-dimensional $\mathrm{O}(2)$ QNLSM
at zero temperature, $\xi$
is the inverse of the energy gap
(in units in which the velocity $c^{\,}_{\mathfrak{a}}$
and the Planck constant $\hbar$ are unity)
to excite a longitudinal mode
[$\sigma$ in the spin-wave parametrization
(\ref{eq:d-dimensional O(N) QNLSM non zero T e})],
i.e., a mode that changes the magnitude of
the staggered magnetization, as opposed to the Goldstone modes
[$\bm{\pi}$ in the spin-wave parametrization
(\ref{eq:d-dimensional O(N) QNLSM non zero T e})]
that merely change the orientation of the staggered magnetization.
From the point of view of the classical XY model
on the cubic lattice in thermodynamic equilibrium
at the inverse temperature $\beta^{\,}_{\mathrm{eff}}$,
$\xi$ is a measure of how far $\beta^{\,}_{\mathrm{eff}}$ is
above the critical value $\beta^{\,}_{\mathrm{eff}, \mathrm{c}}$
below which the classical XY model on the cubic lattice
enters its paramagnetic phase.

\subsection{Vicinity to the phase boundary at $\gamma=0$ and
$\beta^{\,}_{\mathrm{BKT}}\leq\beta<\infty$}

At any non-vanishing temperature,
i.e.,
\begin{subequations}
\label{eq:existence Matsubara level spacing}
\begin{equation}
0<\frac{1}{\beta}\lesssim J,
\label{eq:existence Matsubara level spacing a}
\end{equation}
the Fourier expansions
of the imaginary-time dependencies of the $N-1$ transverse modes
$\bm{\pi}(\tau,\bm{r})\in\mathbb{R}^{N-1}$
defined in Eq.\ (\ref{eq:d-dimensional O(N) QNLSM non zero T e})
in terms of the Matsubara modes
$\bm{\pi}(\varpi^{\,}_{l},\bm{r})\in\mathbb{C}^{N-1}$
with the Matsubara frequencies $\varpi^{\,}_{l}=(2\pi/\beta)\,l$,
where $l\in\mathbb{Z}$, implies the existence of a non-vanishing
Matsubara energy gap
\begin{equation}
\Delta^{\,}_{\mathrm{Mat}}\coloneqq\frac{2\pi}{\beta}>0.
\label{eq:existence Matsubara level spacing b}
\end{equation}
\end{subequations}
Integration in the path integral
(\ref{eq:d-dimensional O(N) QNLSM non zero T a})
over the Matsubara modes
$\bm{\pi}(\varpi^{\,}_{l},\bm{r})$
with non-vanishing Matsubara frequencies
$\varpi^{\,}_{l}\neq0$
can be performed perturbatively.
The outcome,
to lowest order in this expansion when $d=2$ and $N=2$,
and if we impose $\mathrm{O}(2)$ symmetry by demanding that
\begin{equation}
\gamma=0,
\label{eq:imposing O(2) symmetry}  
\end{equation}
is the two-dimensional classical $\mathrm{O}(2)$ NLSM
\cite{Chakravarty89}
\begin{subequations}
\begin{widetext}
\label{eq:2-dimensional O(2) QNLSM non zero T final}
\begin{align}
&
Z^{\mathrm{eff}}_{\mathrm{O}(2)}
(\beta,\rho^{\,}_{\mathfrak{a}},c^{\,}_{\mathfrak{a}})=
\left[
\prod_{\bm{r}\in\mathbb{R}^{2}}^{|\bm{r}|\geq\mathfrak{a}}\,
\int
\mathrm{d}^{2}\bm{n}(\bm{r})\,
\delta\bm{\big(}\bm{n}^{2}(\bm{r})-1\bm{\big)}
\right]
\exp
\left(
-
\frac{
	S^{\mathrm{eff}}_{\mathrm{cla}}(\beta,\rho^{\,}_{\mathfrak{a}},c^{\,}_{\mathfrak{a}})
}
{
	\hbar
}
\right),
\label{eq:2-dimensional O(2) QNLSM non zero T a final}
\\
&
S^{\mathrm{eff}}_{\mathrm{cla}}
(\beta,\rho^{\,}_{\mathfrak{a}},c^{\,}_{\mathfrak{a}})=
\frac{1}{2\,t^{\,}_{\mathfrak{a}}(\beta,\rho^{\,}_{\mathfrak{a}},c^{\,}_{\mathfrak{a}})}
\int\limits_{|\bm{r}|\geq\mathfrak{a}}\mathrm{d}^{2}\bm{r}\,
\sum_{a=1}^{2}
\left(
\frac{\partial\bm{n}}{\partial r^{\,}_{a}}
\right)^{2}\!\!(\bm{r}),
\label{eq:2-dimensional O(2) QNLSM non zero T b final}
\end{align}
\end{widetext}
where the dimensionless coupling
$t^{\,}_{\mathfrak{a}}(\beta)$
is given by
\begin{equation}
\frac{1}{t^{\,}_{\mathfrak{a}}(\beta,\rho^{\,}_{\mathfrak{a}},c^{\,}_{\mathfrak{a}})}=
\beta
\left[
\rho^{\,}_{\mathfrak{a}}
+
\mathcal{O}(\beta^{-2})
\right].
\label{eq:2-dimensional O(2) QNLSM non zero T c final}
\end{equation}
Whereas the parametrization
(\ref{eq:d-dimensional O(N) QNLSM non zero T e})
is limited to the half circle when $N=2$, the parametrization
\begin{equation}
\bm{n}=\begin{pmatrix}\cos\phi\\\sin\phi\end{pmatrix}
\label{eq:2-dimensional O(2) QNLSM non zero T d final}
\end{equation}
covers the full circle and yields
\begin{equation}
\sum_{a=1}^{2}
\left(
\frac{\partial\bm{n}}{\partial r^{\,}_{a}}
\right)^{2}\!\!(\bm{r})=
(\bm{\nabla}\phi)^{2}(\bm{r}).
\label{eq:2-dimensional O(2) QNLSM non zero T e final}
\end{equation}
\end{subequations}
This parametrization,
as opposed to that in Eq.\ (\ref{eq:d-dimensional O(N) QNLSM non zero T e}),
preserves the full $\mathrm{O}(2)$ symmetry.

We relax the $\mathrm{O}(2)$ symmetry
by replacing condition
(\ref{eq:imposing O(2) symmetry})
with condition
\begin{subequations}
\begin{equation}
|\gamma|\ll1.
\end{equation}
The anisotropic exchange coupling
\begin{equation}
-
\gamma\,J\,
\left(
\hat{S}^{x}_{j}\,\hat{S}^{x}_{j'}
-
\hat{S}^{y}_{j}\,\hat{S}^{y}_{j'}
\right)
\end{equation}
for any pair of nearest neighbor sites $j$ and $j'$
generates the nonlinear perturbation
\begin{equation}
+
\gamma\,
J\,
S^{2}\,
\cos(2\phi)
\end{equation}
\end{subequations}
to the two-dimensional classical $\mathrm{O}(2)$ NLSM
(\ref{eq:2-dimensional O(2) QNLSM non zero T final}).
This perturbation breaks the $\mathrm{O}(2)$ symmetry
down to its $\mathbb{Z}^{\,}_{2}\times\mathbb{Z}^{\,}_{2}$ subgroup 
generated by the two transformations
\begin{align}
\phi \mapsto -\phi,
\qquad
\phi \mapsto -\phi+\pi,
\end{align}
under both of which $\cos(2\phi)$ is invariant
\footnote{
On the one hand,
all harmonics $\cos(2q\phi)$ with $q=1,2,\cdots$
are allowed by the internal symmetry
$\mathbb{Z}^{\,}_{2}\times\mathbb{Z}^{\,}_{2}$
and by the symmetry under reversal of time, as is emphasized in Ref.\
\cite{Prakash24}.
However, the higher harmonics ($q=2,3,\cdots$) are not present in
Hamiltonian
(\ref{eq:def anisotropic quantum spin-1/2 XY Lambda})
and need not be included as bare interactions in the
Lagrangian density (\ref{eq:effective field theory at T>0 c}),
as our sole purpose is to study the phase diagram of
Hamiltonian
(\ref{eq:def anisotropic quantum spin-1/2 XY Lambda})
(see also footnote \ref{foonotnote:2}).
On the other hand, since the harmonics $\cos(2q\phi)$ are
not invariant under the O(2) transformation, $\phi\to\phi+\mathrm{const}$.,
they all have to vanish when the only O(2)-symmetry breaking
parameter $\gamma$ is turned off in Hamiltonian
(\ref{eq:def anisotropic quantum spin-1/2 XY Lambda}).
         }.
Correspondingly, at any non-vanishing temperature
(\ref{eq:existence Matsubara level spacing a}),
we thus trade the partition function of Hamiltonian
(\ref{eq:def anisotropic quantum spin-1/2 XY Lambda})
with $\Lambda$ a square lattice
for the classical finite-temperature field theory with the partition function
\begin{subequations}
\label{eq:effective field theory at T>0}
\begin{align}
&
Z(t^{\,}_{\mathfrak{a}},\gamma^{\,}_{\mathfrak{a}})\coloneqq
\int\!\mathcal{D}[\varphi]\,
\delta
\left(
\bm{\nabla}\wedge\bm{\nabla}\varphi
\right)
\nonumber\\
&
\hphantom{
Z(t^{\,}_{\mathfrak{a}},\gamma^{\,}_{\mathfrak{a}})\coloneqq
}
\times
\exp\!\left(\,
\int\limits_{\mathbb{R}^{2}}\mathrm{d}^{2}\bm{r}
\left(
\mathcal{L}^{\,}_{\varphi}
+
\mathcal{L}^{\,}_{\mathrm{ani}}
\right)
\right),
\label{eq:effective field theory at T>0 a}
\\
&
\mathcal{L}^{\,}_{\varphi}\coloneqq
\frac{1}{2\,t^{\,}_{\mathfrak{a}}}\,
(\bm{\nabla}\varphi)^{2},
\label{eq:effective field theory at T>0 b}
\\
&
\mathcal{L}^{\,}_\mathrm{ani}\coloneqq
\gamma^{\,}_{\mathfrak{a}}\,
\cos(2\varphi),
\label{eq:effective field theory at T>0 c}
\end{align}
\end{subequations}
where 
(i) the bare value of the dimensionless coupling
$t^{\,}_{\mathfrak{a}}$ is $(\beta\,J\,S^{2})^{-1}$,
(ii) the bare value of the dimensionful coupling
$\gamma^{\,}_{\mathfrak{a}}$
is
$\gamma\,\beta\,J\,S^{2}/\mathfrak{a}^{2}$,
and (iii) we impose through the delta function
on the right-hand side of Eq.\
(\ref{eq:effective field theory at T>0 a})
the condition that $\bm{\nabla}\varphi$
is rotation free, i.e., we exclude the possibility,
present in $\phi$ defined by Eq.\
(\ref{eq:2-dimensional O(2) QNLSM non zero T d final}),
that $\varphi$ is not single-valued everywhere and
thus supports vortices.

With the help of the Green's function defined by
\begin{equation}
-\kappa^{-1}\,\Delta\,G(\bm{r})=\delta(\bm{r})
\
\Longleftrightarrow
\
G(\bm{r})=
-\frac{\kappa}{2\pi}\,\ln\left|\frac{\bm{r}}{\mathfrak{a}}\right|
\end{equation}
for any non-vanishing real-valued number $\kappa$,
we find that the two-point spin correlation functions
show algebraic decay with an exponent proportional to the temperature,
\begin{equation}
\langle
\bm{n}(\bm{r})\cdot\bm{n}(\bm{0})\rangle_{\gamma=0}\sim
\langle
e^{+\mathrm{i}\varphi(\bm{r})}
e^{-\mathrm{i}\varphi(\bm{0})}
\rangle^{\,}_{\gamma=0}=
\left|\frac{\mathfrak{a}}{\bm{r}}\right|^{+\frac{t^{\,}_{\mathfrak{a}}}{2\pi}},
\label{eq:algebraic decay}
\end{equation}
when $\gamma=0$.
There follows the scaling dimension
\begin{subequations}
\begin{equation}
\Delta^{\,}_{\cos(2\varphi)}=
\frac{t^{\,}_{\mathfrak{a}}}{\pi}
\end{equation}
of the perturbation
$\cos(2\varphi)$
and the renormalization-group flow
\footnote{\label{foonotnote:2}
All higher harmonics $\cos(2q\phi)$ with $q=2,3,\cdots$
are generated by the products of $\cos(2\phi)$
under the renormalization group flow,
but their scaling dimension $q^{2}\,\Delta^{\,}_{\cos(2\varphi)}$
is larger than that of $\cos(2\phi)$ when $\gamma=0$. Hence,
their generation under the renormalization group flow induced
by any small value of $\gamma$
does not change the phase diagram, as these perturbations
are not to be treated as independent perturbations
for Hamiltonian
(\ref{eq:def anisotropic quantum spin-1/2 XY Lambda}).
To treat the higher harmonic
$\cos(2q\phi)$
as an independent perturbation to the
$\mathrm{O}(2)$ symmetric point $\gamma=0$,
one must perturb
Hamiltonian
(\ref{eq:def anisotropic quantum spin-1/2 XY Lambda})
with local terms of order $2q$ in the quantum
spin-{$\frac{1}{2}$} operators~\cite{Prakash24}.
         }
\begin{equation}
\frac{\mathrm{d}\gamma(\ell)}{\mathrm{d}\ell}=
\left(2-\frac{t^{\,}_{\mathfrak{a}}}{\pi}\right)\gamma(\ell)
\end{equation}
with the initial value
\begin{equation}
\gamma(\ell=0)\equiv
\gamma\sim
\frac{\mathfrak{a}^{2}\,\gamma^{\,}_{\mathfrak{a}}}{\beta\,J\,S^{2}},
\end{equation}
\end{subequations}
that is obeyed by the running dimensionless coupling
$\gamma(\ell)$
under the rescaling
$\mathfrak{a}\to\exp(\mathrm{d}\ell)\,\mathfrak{a}$
of the short-distance cutoff.
The correlation length
\begin{subequations}
\label{eq:xi_gamma}
\begin{equation}
\xi\coloneqq
\mathfrak{a}\,
e^{\ell}
\label{eq:xi_gamma a}
\end{equation}
at which $|\gamma(\ell)|$
is of order unity is
\begin{equation}
\xi(\gamma,\beta)\sim
\mathfrak{a}\,
|\gamma|^{-\left(2-\frac{t^{\,}_{\mathfrak{a}}}{\pi}\right)^{-1}}=
\mathfrak{a}\,
|\gamma|^{-\frac{\pi}{2\pi-t^{\,}_{\mathfrak{a}}}}
\label{eq:xi_gamma b}
\end{equation}
\end{subequations}
in the regime $0<t^{\,}_{\mathfrak{a}}<2\pi$
when $\mathcal{L}^{\,}_{\mathrm{ani}}$ is a relevant perturbation
to  $\mathcal{L}^{\,}_{\varphi}$.
Since the order parameter
$\bm{m}^{\,}_{\mathrm{sta}}\sim\bm{n}\sim e^{\mathrm{i}\varphi}$
has the scaling dimension $t^{\,}_{\mathfrak{a}}/4\pi$,
we expect that the dependence of
$|\bm{m}^{\,}_{\mathrm{sta}}|$
on $|\gamma|$ that is inherited from
$\xi(\gamma,\beta)$ should be given by
\begin{equation}
|\bm{m}^{\,}_{\mathrm{sta}}(\gamma,\beta)|\sim
\left(
\frac{\mathfrak{a}}{\xi(\gamma,\beta)}
\right)^{
+\frac{t^{\,}_{\mathfrak{a}}}{4\pi}
}\sim
|\gamma|^{
+\frac{t^{\,}_{\mathfrak{a}}}{4(2\pi-t^{\,}_{\mathfrak{a}})}
}
\label{eq:|m| for small gamma}
\end{equation} 
for $0<|\gamma|\ll1$ and $0<t^{\,}_{\mathfrak{a}}<2\pi$.
We emphasize again
that $\bm{m}_\mathrm{sta}\parallel(\pm1,0)$ when $\gamma<0$
and $\bm{m}_\mathrm{sta}\parallel(0,\pm1)$ when $\gamma>0$.
We have thus found that the staggered magnetization exhibits a power-law
dependence on the exchange anisotropy $\gamma$
with the scaling exponent proportional to the temperature.
This feature is in line with the finite-temperature continuous phase transition
line $\gamma=0$ ending at the first-order transition at zero temperature
in the phase diagram, Fig.~\ref{fig:phase diagrams d=1 and d=2}(b).

\begin{figure}[t!]
\includegraphics[width=0.8\columnwidth]{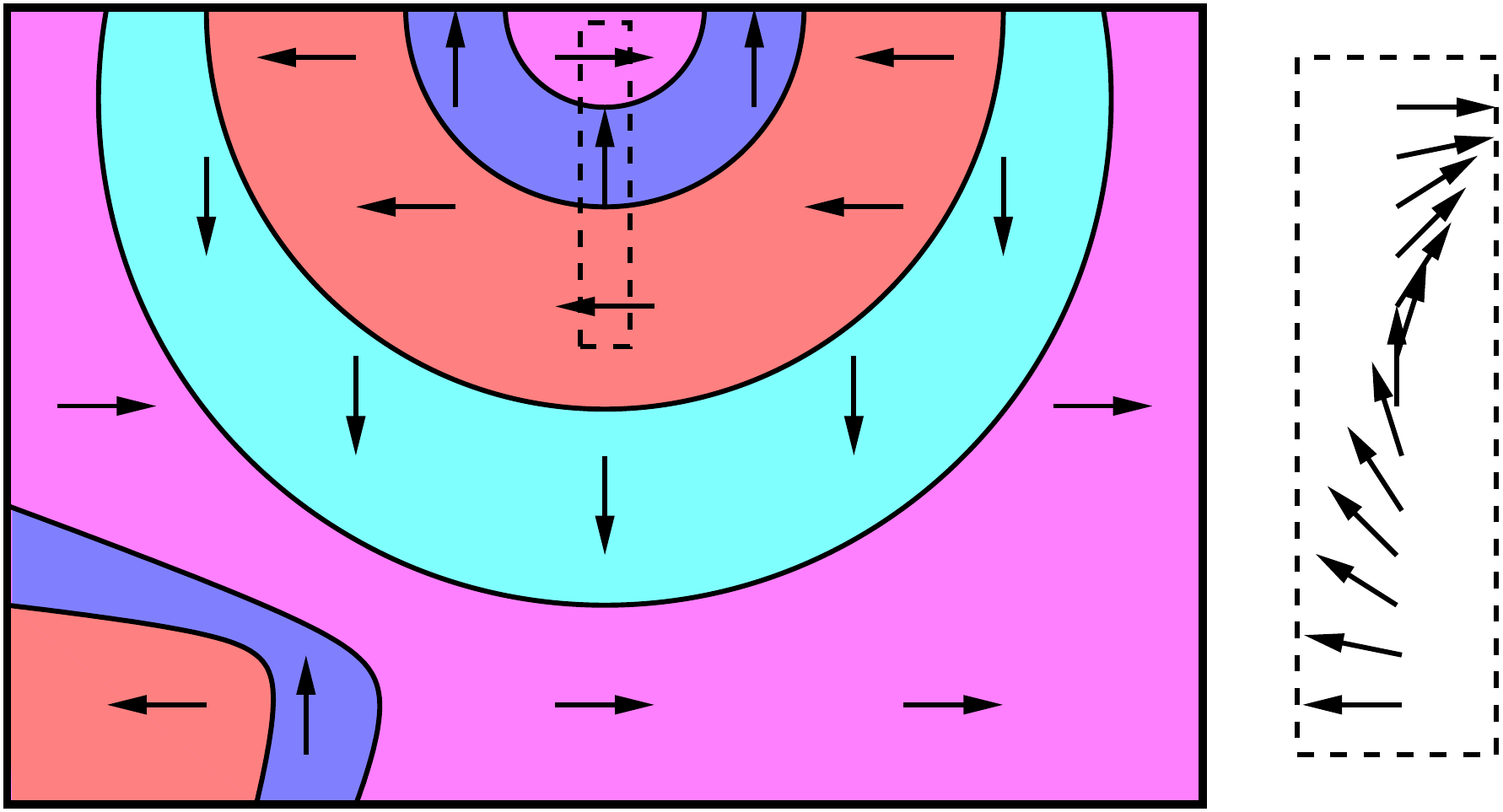}
\caption{(Color online)
Peierls argument reinterpreted as the deconfinement of
(one-dimensional) defects nucleating a competing order.
The order parameter is represented by a unit vector,
whose direction is a smooth function of space; as is shown
in the inset (enclosed by the rectangle with dashed boundaries)
along a one-dimensional cut of two-dimensional
space across a domain wall (colored in blue)
separating a domain with order parameter $(+1,0)$ (colored in magenta)
and a domain with order parameter $(-1,0)$ (colored in red).
}
\label{fig:Peierls argument revisited}
\end{figure}

We can adapt Peierls argument \cite{Peierls36}
in support of a phase transition in the classical
two-dimensional Ising model to argue that $\gamma$
drives a continuous phase transition through the proliferation of
one-dimensional domain walls upon approaching the critical value
$\gamma=0$ as is shown in Fig.\
\ref{fig:Peierls argument revisited}.
To see this we make two observations:
\begin{itemize}
\item The minimum of the perturbation
\begin{subequations}  
\begin{equation}
\mathcal{L}^{\,}_{\mathrm{ani}}\coloneqq
\gamma^{\,}_{\mathfrak{a}}\,
\cos(2\varphi),
\qquad
\gamma^{\,}_{\mathfrak{a}}<0,
\end{equation}
is
\begin{equation}
\varphi=0,\pi,
\end{equation}
in which case it is the N\'eel${}^{\,}_{x}$ phase that is selected
by the flow to strong coupling of $\mathcal{L}^{\,}_{\mathrm{ani}}$
since
\begin{equation}
(\cos\varphi,\sin\varphi)=(+1,0), (-1,0).
\end{equation}
\end{subequations} 
\item The minimum of the perturbation
\begin{subequations}  
\begin{equation}
\mathcal{L}^{\,}_{\mathrm{ani}}\coloneqq
\gamma^{\,}_{\mathfrak{a}}\,
\cos(2\varphi),
\qquad
\gamma^{\,}_{\mathfrak{a}}>0,
\end{equation}
is
\begin{equation}
\varphi=-\frac{\pi}{2},+\frac{\pi}{2},
\end{equation}
in which case the N\'eel${}^{\,}_{y}$ phase is selected
by the flow to strong coupling of $\mathcal{L}^{\,}_{\mathrm{ani}}$
since
\begin{equation}
(\cos\varphi,\sin\varphi)=(0,-1), (0,+1).
\end{equation}
\end{subequations} 
\end{itemize}
When $\gamma\uparrow0$
at any fixed temperature $0<T<T^{\,}_{\mathrm{BKT}}$,
the correlation length diverges. Within a correlation length,
N\'eel${}^{\,}_{x}$ order with one orientation is found in
two-dimensional space, see Fig.\
\ref{fig:Peierls argument revisited}.
Across the boundary (a one-dimensional domain wall)
between two such domains defined by a smooth change
of $\varphi$ from the values $0$ to $\pi$ or $\pi$ to $0$,
$\varphi$ must take either one of the values
$-\pi/2,+\pi/2$
at the boundary, i.e.,
the boundary nucleate the N\'eel${}^{\,}_{y}$ order, see Fig.\
\ref{fig:Peierls argument revisited}.
The same argument applies except for the interchange of the
N\'eel${}^{\,}_{x}$ and N\'eel${}^{\,}_{y}$ order
when $\gamma\downarrow0$.
The line $\gamma=0$
realizes at each temperature $0<T<T^{\,}_{\mathrm{BKT}}$
a continuous phase boundary separating two ordered phase that
are not related by spontaneous symmetry breaking.
Any one-dimensional domain wall
in one ordered phase nucleates the order on the other side of the
continuous phase boundary. For this reason,
we call the line $\gamma=0$
with the temperature $0<T<T^{\,}_{\mathrm{BKT}}$
a line of classical deconfined critical points.

The discussion based on the effective theory
(\ref{eq:effective field theory at T>0})
is valid at low temperatures
until the critical line $\gamma=0$
is terminated by the tricritical point at the
BKT transition temperature $T=T^{\,}_{\mathrm{BKT}}$.
As is well known \cite{JKKN77},
the BKT transition of the
two-dimensional classical XY model
or the O(2) NLSM
(\ref{eq:2-dimensional O(2) QNLSM non zero T final})
can be described by the sine-Gordon model,
\begin{subequations}
\label{eq:effective field theory at T>0 vortices}
\begin{align}
&
Z^{\,}_{\mathrm{BKT}}
(t^{\,}_{\mathfrak{a}},y^{\,}_{\mathfrak{a}})\coloneqq
\int\mathcal{D}[\theta]
\exp\left(-
\int\limits_{\mathbb{R}^{2}}\mathrm{d}^{2}\bm{r}\,
\left(
\mathcal{L}^{\,}_{\theta}
+
\mathcal{L}^{\,}_{\mathrm{vrt}}
\right)\right),
\label{eq:effective field theory at T>0 vortices a}
\\
&
\mathcal{L}^{\,}_{\theta}\coloneqq
\frac{t^{\,}_{\mathfrak{a}}}{2}\,
(\bm{\nabla}\theta)^{2},
\label{eq:effective field theory at T>0 vortices b}
\\
&
\mathcal{L}^{\,}_{\mathrm{vrt}}\coloneqq
y^{\,}_{\mathfrak{a}}\,
\cos(2\pi\theta),
\label{eq:effective field theory at T>0 vortices c}
\end{align}
\end{subequations}
where the bare value of the dimensionless coupling
$t^{\,}_{\mathfrak{a}}$ is again $(\beta\,J\,S^{2})^{-1}$ and
the dimensionful coupling $y^{\,}_{\mathfrak{a}}$
is related to the dimensionless fugacity $Y$ of charge one vortices by
$Y\sim|\mathfrak{a}^{2}\,y^{\,}_{\mathfrak{a}}|\ll1$.

Since the scaling dimension of the operator $\cos(2\pi\theta)$
in the free theory $\mathcal{L}^{\,}_{\theta}$ is
$\Delta_{\cos(2\pi\theta)}=\pi/t^{\,}_{\mathfrak{a}}$,
the perturbation (\ref{eq:effective field theory at T>0 vortices c})
becomes relevant and vortices proliferate to destroy
the critical spin correlations (\ref{eq:algebraic decay})
when $t^{\,}_{\mathfrak{a}}>\frac{\pi}{2}$, i.e.,
\begin{equation}
T>T^{\,}_{\mathrm{BKT}}=\frac{\pi}{2}\,J\,S^{2}.
\end{equation}
This formula for the BKT transition temperature or the tricritical point
is valid in the limit $|\mathfrak{a}^{2}\,y^{\,}_{\mathfrak{a}}|\ll1$.
To account for a small but non-vanishing fugacity $Y$ of vortices,
needed are the linearized one-loop
BKT renormalization-group flows
\begin{subequations}
\label{eq:BKT RG flows}
\begin{align}
&
\frac{\mathrm{d}Y(\ell)}{\mathrm{d}\ell}=
\left(2-\frac{\pi}{t(\ell)}\right)Y(\ell)
\nonumber\\
&\hphantom{\frac{\mathrm{d}Y(\ell)}{\mathrm{d}\ell}}
=
2X(\ell)\,Y(\ell) +\mathcal{O}(X^{2}\,Y),
\\
&
\frac{\mathrm{d}X(\ell)}{\mathrm{d}\ell}=
A\,Y^{2}(\ell),
\qquad
X(\ell)\coloneqq
\frac{2}{\pi}\,t(\ell)
-
1,
\end{align}
with the initial values
\begin{align}
&
Y(\ell=0)\equiv Y\sim|\mathfrak{a}^{2}\,y^{\,}_{\mathfrak{a}}|,
\\
&
X(\ell=0)\equiv
X\coloneqq
\frac{2}{\pi}\,t^{\,}_{\mathfrak{a}}
-
1,
\end{align}
\end{subequations}
that are obeyed by the running dimensionless couplings
$Y(\ell)$ and $X(\ell)$
under the rescaling $\mathfrak{a}\to\exp(\mathrm{d}\ell)\,\mathfrak{a}$
of the short-distance cutoff.
Here, $A$ is a regularization-dependent positive constant.
The BKT transition temperature is then determined from the separatrix
$\sqrt{2}\,X=-\sqrt{A}\,Y$ 
of the renormalization-group flows (\ref{eq:BKT RG flows}).
Integration of the scaling equations (\ref{eq:BKT RG flows})
gives the estimate
\begin{equation}
\xi^{\,}_{\mathrm{BKT}}(t)\sim\mathfrak{a}\,e^{\alpha/\sqrt{t}},
\qquad  
t\coloneqq
\frac{T-T^{\,}_{\mathrm{BKT}}}{T^{\,}_{\mathrm{BKT}}},
\label{eq:xi_BKT}
\end{equation}
where $\alpha$ is a positive constant,
for how the correlation length diverges upon approaching $T=T^{\,}_{\mathrm{BKT}}$
from above.

The two phase boundaries that connect the
tricritical point to the N\'eel${}^{\,}_{x}$
and N\'eel${}^{\,}_{y}$ Ising fixed points
[Fig.\ \ref{fig:phase diagrams d=1 and d=2}(b)]
can be located in the close vicinity of
the tricritical point
by equating the correlation length $\xi^{\,}_{\mathrm{BKT}}$ in Eq.\ 
(\ref{eq:xi_BKT})
with the correlation length
$\xi(\gamma,\beta)$
at $t^{\,}_{\mathfrak{a}}=\frac{\pi}{2}$
in Eq.\ (\ref{eq:xi_gamma}).
This gives the relation
\begin{equation}
\gamma=\pm
\exp
\left(
-
\frac{3\alpha}{2}\,
\sqrt{\frac{T^{\,}_{\mathrm{BKT}}}{T-T^{\,}_{\mathrm{BKT}}}}
\right),
\end{equation}
for the phase boundary emerging from
the tricritical point in
Fig.\ \ref{fig:phase diagrams d=1 and d=2}(b)
for which the running coupling constants
$\gamma(\ell)$
and
$Y(\ell)$
are of order one.

\begin{figure}[t!]
\includegraphics[width=1\columnwidth]{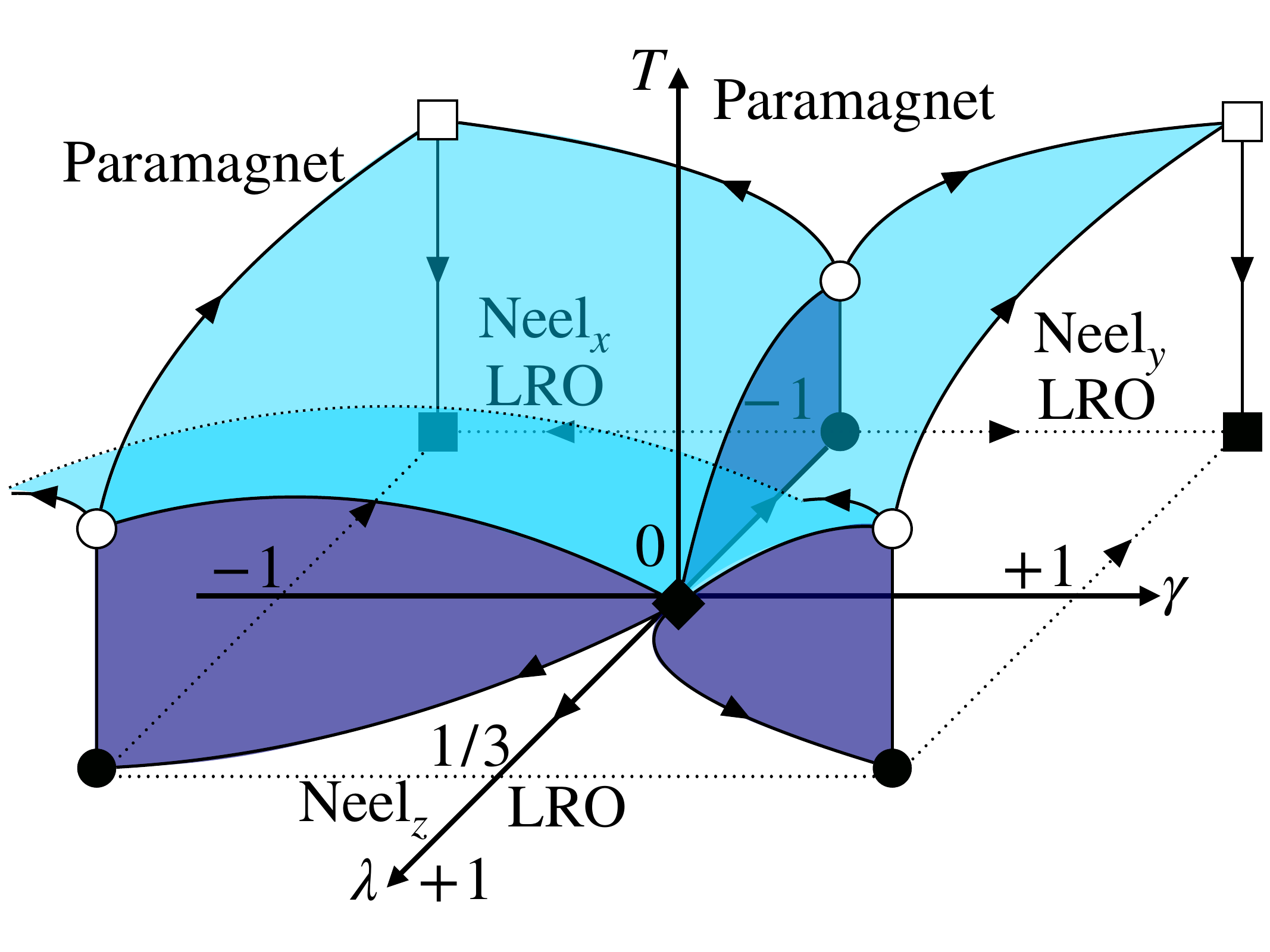}
\caption{(Color online)
Phase diagram of the two-dimensional anisotropic antiferromagnetic 
quantum spin-{$\frac{1}{2}$} XYZ model defined by Hamiltonian
(\ref{eq:def anisotropic quantum spin-1/2 XYZ Lambda})
with $\Lambda$ a square lattice. The set of symbols common to
Fig.\ \ref{fig:phase diagrams d=1 and d=2}(b)
have the same meaning. 
The solid circle from Fig.\
\ref{fig:phase diagrams d=1 and d=2}(b)
is now located at $\lambda=-1$, $\gamma=0$, and $T=0$.
Two new solid circles are located at
$\lambda=1/3$, $\gamma=\pm1$, and $T=0$.
A new fixed point denoted by a diamond is present when
$\lambda=0$, $\gamma=0$, and $T=0$.
This fixed point is unstable to any deviation away from
$\lambda=0$, $\gamma=0$, and $T=0$. It realizes an
antiferromagnetic long-ranged ordered phase with full $\mathrm{SU}(2)$
symmetry in that the staggered magnetization is non-vanishing
and can point in any of the three directions in spin-{$\frac{1}{2}$} space.
There are three surfaces colored in dark blue that meet
at $\lambda=0$, $\gamma=0$, and $T=0$. Each one of these dark blue
surfaces separates two of the three Ising phases
N\'eel${}^{\,}_{x}$, N\'eel${}^{\,}_{y}$, and N\'eel${}^{\,}_{z}$
phases. A paramagnetic phase sits on top of the three Ising phases
and is separated from them by the light blue surfaces.
Any two of the light blue surfaces intersect along boundaries
of the dark blue surfaces.
The arrows represent the renormalization group flows.
}
\label{fig:phase diagrams d=2, XYZ model}.
\end{figure}

\section{Discussion}
\label{sec:Discussion}

The deconfined classical criticality studied in this paper is a
robust property of the lattice two-spin-interaction Hamiltonian
(\ref{eq:def anisotropic quantum spin-1/2 XY Lambda}).
It is not a stable property of a generic local lattice Hamiltonian
sharing the same symmetries as Hamiltonian
(\ref{eq:def anisotropic quantum spin-1/2 XY Lambda}),
as is emphasized in Ref.\ \cite{Prakash24}. Indeed,
such a generic lattice Hamiltonian can have symmetry-broken
phases at finite temperatures that are distinct from the
N\'eel${}^{\,}_{x}$
and
N\'eel${}^{\,}_{y}$
phases \cite{Prakash24}.
For such a generic lattice Hamiltonian,
fine-tuning of at least two
coupling constants is necessary to observe deconfined classical
multicriticality \cite{Prakash24}. In this regard,
the deconfined classical transition
between the
N\'eel${}^{\,}_{x}$
and
N\'eel${}^{\,}_{y}$
phases is different from that taking place at zero temperature
for the quantum spin-{$\frac{1}{2}$} chains studied in Ref.\ \cite{Mudry19}.
In the quantum spin-{$\frac{1}{2}$} chains studied in Ref.\ \cite{Mudry19},
there is only one relevant operator at deconfined quantum
critical points separating the zero-temperature
N\'eel${}^{\,}_{x}$
and
N\'eel${}^{\,}_{y}$
phases.
Here, on the other hand,
the number of relevant operators
at the finite-temperature continuous phase transition between
the
N\'eel${}^{\,}_{x}$
and
N\'eel${}^{\,}_{y}$
phases is a finite number larger than one that increases with lowering
the temperature and eventually
diverges as the temperature approaches zero.
One may thus interpret the discontinuous nature of the
quantum phase transition between the
N\'eel${}^{\,}_{x}$
and
N\'eel${}^{\,}_{y}$
phases at zero temperature
for the lattice two-spin-interaction Hamiltonian
(\ref{eq:def anisotropic quantum spin-1/2 XY Lambda})
as the divergence in the number of symmetry-allowed
relevant perturbations at the classical continuous phase transition
between the
N\'eel${}^{\,}_{x}$
and
N\'eel${}^{\,}_{y}$
phases as the temperature approaches zero.

The same conclusions reached in
Secs.\
\ref{sec:Discontinuous quantum phase transition}
and
\ref{sec:Deconfined classical criticality when ...}
that are summarized in Fig.\
\ref{fig:phase diagrams d=1 and d=2}(b)
apply to the anisotropic quantum spin-{$\frac{1}{2}$} YZ and ZX 
Hamiltonians, which are defined by the cyclic permutations
of the operators $\hat{S}^{x}_{j}$, $\hat{S}^{y}_{j}$,
and
$\hat{S}^{z}_{j}\equiv-\mathrm{i}\hat{S}^{x}_{j}\,\hat{S}^{y}_{j}$
in Eq.\ (\ref{eq:def anisotropic quantum spin-1/2 XY Lambda}).
Hence, if we define the family
\begin{equation}
\hat{H}^{\,}_{\Lambda}(\gamma,\lambda)\coloneqq
(1-\lambda)\,
\hat{H}^{\,}_{\Lambda}(\gamma)
+
(1+\lambda)\,
J\,
\sum_{\langle jj'\rangle}
\hat{S}^{z}_{j}\,
\hat{S}^{z}_{j'}
\label{eq:def anisotropic quantum spin-1/2 XYZ Lambda}
\end{equation}
labeled by $-1\leq\gamma\leq+1$ and $-1\leq\lambda\leq+1$
of two-dimensional anisotropic antiferromagnetic quantum spin-{$\frac{1}{2}$} XYZ 
Hamiltonians, we then have the phase diagram shown in
Fig.\
\ref{fig:phase diagrams d=2, XYZ model},
as we now explain.

At the point
\begin{equation}
\gamma=0,
\qquad
\lambda=0,
\label{eq:def Heisenberg point}
\end{equation}
Hamiltonian (\ref{eq:def anisotropic quantum spin-1/2 XYZ Lambda})
is the nearest-neighbor antiferromagnetic quantum spin-{$\frac{1}{2}$} Heisenberg
Hamiltonian on the square lattice
with the global symmetry group $\mathrm{SU}(2)$.
Its ground-state expectation value
supports a non-vanishing staggered magnetization 
that can point in any direction in spin space
(see Ref.\ \cite{Manousakis91} and references therein
for the numerical evidences supporting this conjecture)
and whose magnitude reaches its absolute minimum
as a function of $-1\leq\lambda\leq+1$,
along the segment
\begin{equation}
T=0,
\qquad
\gamma=0,
\qquad
-1\leq\lambda\leq+1,
\label{eq:line T=gamma=0 0<lambda}
\end{equation} 
as quantum fluctuations are the strongest at this point
(\ref{eq:def Heisenberg point}).
Exact results state that the XY and N\'eel${}^{\,}_{z}$
long-range orders, known to be present in the
ground states corresponding to the XY limit $\lambda=-1$
and the N\'eel${}^{\,}_{z}$ limit $\lambda=+1$,
survive along the two segments
\begin{equation}
T=0,
\qquad
\gamma=0,
\qquad
-1\leq\lambda<-0.87
\end{equation}
and 
\begin{equation}
T=0,
\qquad
\gamma=0,
\qquad
0.78<\lambda\leq1,
\end{equation}
respectively
\cite{Frohlich77,Kubo88,Ozeki89}.
Furthermore, exact-diagonalization and Monte-Carlo simulations
\cite{Barnes89,Ding90b,Ding92b,Aplesnin98,lin01,Roscilde03,Cuccoli03}
support the conjecture that
(i) the XY long-range order
extends to the segment
\begin{equation}
T=0,
\qquad
\gamma=0,
\qquad
-0.87\leq\lambda<0,  
\end{equation}
and (ii) the N\'eel${}^{\,}_{z}$ long-range order
extends to the segment
\begin{equation}
T=0,
\qquad
\gamma=0,
\qquad
0<\lambda\leq0.78.
\end{equation}

In summary, the point (\ref{eq:def Heisenberg point})
is an unstable fixed point with respect to any
(1)
non-vanishing temperature $T>0$
that drives this long-range phase to the paramagnetic phase
through the renormalized classical scaling regime \cite{Chakravarty89},
(2) deviation $\lambda>0$ that drives this long-range ordered phase to 
the N\'eel${}^{\,}_{z}$ phase
\cite{Barnes89,Ding90b,Aplesnin98,lin01,Roscilde03,Cuccoli03},
(3) deviation $\lambda<0$ that drives this long-range ordered phase to 
the XY phase
\cite{Barnes89,Ding90b,Ding92b,lin01,Roscilde03,Cuccoli03},
and
(4) deviation $\gamma\neq0$ that drives this long-range ordered phase to one
of the N\'eel${}^{\,}_{x}$ or N\'eel${}^{\,}_{y}$ phases.

According to Fig.\ \ref{fig:phase diagrams d=2, XYZ model},
any point along the segment
\begin{equation}
T=0,
\qquad
\gamma=0,
\qquad
-1\leq\lambda\leq0,
\label{eq:line T=gamma=0 0<lambda<1/2}
\end{equation}
realizes a quantum first-order phase transition
when perturbed by $\gamma\neq0$. Its $\mathrm{SU}(2)$-symmetric end point
(\ref{eq:def Heisenberg point})
also realizes a first-order quantum phase transition
when perturbed by $\lambda\neq0$,
as the direction of the staggered magnetization jumps from being directed along
the $z$-axis to being in the $x$-$y$ plane in spin space
as $\lambda\neq0$ changes from positive to negative values.

Finally, we conjecture the existence of three critical
planes at finite temperatures that belong
to the universality class of {$\mathsf{c}=1$} CFT
and generalize the line of critical points
of the XY model for
$(\gamma,\lambda)=(0,-1)$ and $0<T<T_\mathrm{BKT}$.
Any point within the planar phase boundary
\begin{equation}
0<T\leq T^{\,}_{\mathrm{BKT}}(\lambda),
\qquad
\gamma=0,
\qquad
-1\leq\lambda\leq0,
\end{equation}
realizes a deconfined classical criticality that separates
(a)
the N\'eel${}^{\,}_{x}$ and N\'eel${}^{\,}_{y}$ phases.
The same is true for the two planar phase boundaries,
\footnote{The second condition in Eq.\
(\ref{eq:planar phase boundaries (b) and (c)})
is obtained by equating the coefficients multiplying
$\hat{S}^{x}_{j}\,\hat{S}^{x}_{j'}$,
$\hat{S}^{y}_{j}\,\hat{S}^{y}_{j'}$,c
and
$\hat{S}^{z}_{j}\,\hat{S}^{z}_{j'}$
in Hamiltonian
(\ref{eq:def anisotropic quantum spin-1/2 XYZ Lambda}).
}
\begin{equation}
0<T\leq T^{\,}_{\mathrm{BKT}}(\gamma,\lambda),
\qquad
\gamma=\pm \frac{2\lambda}{1-\lambda},
\label{eq:planar phase boundaries (b) and (c)}
\end{equation}
separating at non-vanishing temperatures 
(b) the N\'eel${}^{\,}_{y}$ and N\'eel${}^{\,}_{z}$ phases
and (c) the N\'eel${}^{\,}_{z}$ and N\'eel${}^{\,}_{x}$ phases,
respectively.
All three planar phase boundaries meet at the point
(\ref{eq:def Heisenberg point}).

\begin{acknowledgments}
We would like to thank
{A.}\ Prakash and {N.}\ Jones
for pointing out the multicritical nature of 
the deconfined phase transitions in generic lattice Hamiltonians
and the useful discussions that ensued.
This work is supported by NSF {Grant No.}\  DMR-2022428 (\"O.M.A.),
by DOE {Grant No.}\ DE-FG02-06ER46316 (C.C.),
and by JSPS KAKENHI (Grant No.\ JP19K03680)
and JST CREST (Grant No.\ JPMJCR19T2) {(A.F.)}.
\end{acknowledgments}

\newpage

\bibliography{references}

\begin{thebibliography}{69}%
\makeatletter
\providecommand \@ifxundefined [1]{%
 \@ifx{#1\undefined}
}%
\providecommand \@ifnum [1]{%
 \ifnum #1\expandafter \@firstoftwo
 \else \expandafter \@secondoftwo
 \fi
}%
\providecommand \@ifx [1]{%
 \ifx #1\expandafter \@firstoftwo
 \else \expandafter \@secondoftwo
 \fi
}%
\providecommand \natexlab [1]{#1}%
\providecommand \enquote  [1]{``#1''}%
\providecommand \bibnamefont  [1]{#1}%
\providecommand \bibfnamefont [1]{#1}%
\providecommand \citenamefont [1]{#1}%
\providecommand \href@noop [0]{\@secondoftwo}%
\providecommand \href [0]{\begingroup \@sanitize@url \@href}%
\providecommand \@href[1]{\@@startlink{#1}\@@href}%
\providecommand \@@href[1]{\endgroup#1\@@endlink}%
\providecommand \@sanitize@url [0]{\catcode `\\12\catcode `\$12\catcode
  `\&12\catcode `\#12\catcode `\^12\catcode `\_12\catcode `\%12\relax}%
\providecommand \@@startlink[1]{}%
\providecommand \@@endlink[0]{}%
\providecommand \url  [0]{\begingroup\@sanitize@url \@url }%
\providecommand \@url [1]{\endgroup\@href {#1}{\urlprefix }}%
\providecommand \urlprefix  [0]{URL }%
\providecommand \Eprint [0]{\href }%
\providecommand \doibase [0]{https://doi.org/}%
\providecommand \selectlanguage [0]{\@gobble}%
\providecommand \bibinfo  [0]{\@secondoftwo}%
\providecommand \bibfield  [0]{\@secondoftwo}%
\providecommand \translation [1]{[#1]}%
\providecommand \BibitemOpen [0]{}%
\providecommand \bibitemStop [0]{}%
\providecommand \bibitemNoStop [0]{.\EOS\space}%
\providecommand \EOS [0]{\spacefactor3000\relax}%
\providecommand \BibitemShut  [1]{\csname bibitem#1\endcsname}%
\let\auto@bib@innerbib\@empty
\bibitem [{\citenamefont {Lieb}\ \emph {et~al.}(1961)\citenamefont {Lieb},
  \citenamefont {Schultz},\ and\ \citenamefont {Mattis}}]{Lieb61}%
  \BibitemOpen
  \bibfield  {author} {\bibinfo {author} {\bibfnamefont {E.}~\bibnamefont
  {Lieb}}, \bibinfo {author} {\bibfnamefont {T.}~\bibnamefont {Schultz}},\ and\
  \bibinfo {author} {\bibfnamefont {D.}~\bibnamefont {Mattis}},\ }\bibfield
  {title} {\bibinfo {title} {Two soluble models of an antiferromagnetic
  chain},\ }\href
  {https://doi.org/https://doi.org/10.1016/0003-4916(61)90115-4} {\bibfield
  {journal} {\bibinfo  {journal} {Annals of Physics}\ }\textbf {\bibinfo
  {volume} {16}},\ \bibinfo {pages} {407} (\bibinfo {year} {1961})}\BibitemShut
  {NoStop}%
\bibitem [{\citenamefont {Jordan}\ and\ \citenamefont
  {Wigner}(1928)}]{Jordan28}%
  \BibitemOpen
  \bibfield  {author} {\bibinfo {author} {\bibfnamefont {P.}~\bibnamefont
  {Jordan}}\ and\ \bibinfo {author} {\bibfnamefont {E.}~\bibnamefont
  {Wigner}},\ }\bibfield  {title} {\bibinfo {title} {{\"U}ber das paulische
  {\"a}quivalenzverbot},\ }\href
  {https://api.semanticscholar.org/CorpusID:126400679} {\bibfield  {journal}
  {\bibinfo  {journal} {Zeitschrift f{\"u}r Physik}\ }\textbf {\bibinfo
  {volume} {47}},\ \bibinfo {pages} {631} (\bibinfo {year} {1928})}\BibitemShut
  {NoStop}%
\bibitem [{\citenamefont {{Mudry}}\ \emph {et~al.}(2019)\citenamefont
  {{Mudry}}, \citenamefont {{Furusaki}}, \citenamefont {{Morimoto}},\ and\
  \citenamefont {{Hikihara}}}]{Mudry19}%
  \BibitemOpen
  \bibfield  {author} {\bibinfo {author} {\bibfnamefont {C.}~\bibnamefont
  {{Mudry}}}, \bibinfo {author} {\bibfnamefont {A.}~\bibnamefont {{Furusaki}}},
  \bibinfo {author} {\bibfnamefont {T.}~\bibnamefont {{Morimoto}}},\ and\
  \bibinfo {author} {\bibfnamefont {T.}~\bibnamefont {{Hikihara}}},\ }\bibfield
   {title} {\bibinfo {title} {{Quantum phase transitions beyond Landau-Ginzburg
  theory in one-dimensional space revisited}},\ }\href
  {https://doi.org/10.1103/PhysRevB.99.205153} {\bibfield  {journal} {\bibinfo
  {journal} {\prb}\ }\textbf {\bibinfo {volume} {99}},\ \bibinfo {eid} {205153}
  (\bibinfo {year} {2019})}\BibitemShut {NoStop}%
\bibitem [{\citenamefont {Jiang}\ and\ \citenamefont
  {Motrunich}(2019)}]{Jiang19}%
  \BibitemOpen
  \bibfield  {author} {\bibinfo {author} {\bibfnamefont {S.}~\bibnamefont
  {Jiang}}\ and\ \bibinfo {author} {\bibfnamefont {O.}~\bibnamefont
  {Motrunich}},\ }\bibfield  {title} {\bibinfo {title} {Ising ferromagnet to
  valence bond solid transition in a one-dimensional spin chain: Analogies to
  deconfined quantum critical points},\ }\href
  {https://doi.org/10.1103/PhysRevB.99.075103} {\bibfield  {journal} {\bibinfo
  {journal} {Phys. Rev. B}\ }\textbf {\bibinfo {volume} {99}},\ \bibinfo
  {pages} {075103} (\bibinfo {year} {2019})}\BibitemShut {NoStop}%
\bibitem [{\citenamefont {Prakash}\ and\ \citenamefont {Jones}()}]{Prakash24}%
  \BibitemOpen
  \bibfield  {author} {\bibinfo {author} {\bibfnamefont {A.}~\bibnamefont
  {Prakash}}\ and\ \bibinfo {author} {\bibfnamefont {N.~G.}\ \bibnamefont
  {Jones}},\ }\href {https://arxiv.org/abs/2404.19009} {\bibinfo {title}
  {{Classical origins of Landau-incompatible transitions}}},\ \Eprint
  {https://arxiv.org/abs/2404.19009} {arXiv:2404.19009 [cond-mat.stat-mech]}
  \BibitemShut {NoStop}%
\bibitem [{\citenamefont {Verresen}\ \emph {et~al.}(2017)\citenamefont
  {Verresen}, \citenamefont {Moessner},\ and\ \citenamefont
  {Pollmann}}]{Verresen17}%
  \BibitemOpen
  \bibfield  {author} {\bibinfo {author} {\bibfnamefont {R.}~\bibnamefont
  {Verresen}}, \bibinfo {author} {\bibfnamefont {R.}~\bibnamefont {Moessner}},\
  and\ \bibinfo {author} {\bibfnamefont {F.}~\bibnamefont {Pollmann}},\
  }\bibfield  {title} {\bibinfo {title} {One-dimensional symmetry protected
  topological phases and their transitions},\ }\href
  {https://doi.org/10.1103/PhysRevB.96.165124} {\bibfield  {journal} {\bibinfo
  {journal} {Phys. Rev. B}\ }\textbf {\bibinfo {volume} {96}},\ \bibinfo
  {pages} {165124} (\bibinfo {year} {2017})}\BibitemShut {NoStop}%
\bibitem [{\citenamefont {Aksoy}\ \emph {et~al.}(2024)\citenamefont {Aksoy},
  \citenamefont {Mudry}, \citenamefont {Furusaki},\ and\ \citenamefont
  {Tiwari}}]{Aksoy24}%
  \BibitemOpen
  \bibfield  {author} {\bibinfo {author} {\bibfnamefont {{\"O}.~M.}\
  \bibnamefont {Aksoy}}, \bibinfo {author} {\bibfnamefont {C.}~\bibnamefont
  {Mudry}}, \bibinfo {author} {\bibfnamefont {A.}~\bibnamefont {Furusaki}},\
  and\ \bibinfo {author} {\bibfnamefont {A.}~\bibnamefont {Tiwari}},\
  }\bibfield  {title} {\bibinfo {title} {{Lieb-Schultz-Mattis anomalies and web
  of dualities induced by gauging in quantum spin chains}},\ }\href
  {https://doi.org/10.21468/SciPostPhys.16.1.022} {\bibfield  {journal}
  {\bibinfo  {journal} {SciPost Phys.}\ }\textbf {\bibinfo {volume} {16}},\
  \bibinfo {pages} {022} (\bibinfo {year} {2024})}\BibitemShut {NoStop}%
\bibitem [{\citenamefont {Ising}(1925)}]{Ising25}%
  \BibitemOpen
  \bibfield  {author} {\bibinfo {author} {\bibfnamefont {E.}~\bibnamefont
  {Ising}},\ }\bibfield  {title} {\bibinfo {title} {Beitrag zur theorie des
  ferromagnetismus},\ }\href
  {https://api.semanticscholar.org/CorpusID:122157319} {\bibfield  {journal}
  {\bibinfo  {journal} {Zeitschrift f{\"u}r Physik}\ }\textbf {\bibinfo
  {volume} {31}},\ \bibinfo {pages} {253} (\bibinfo {year} {1925})}\BibitemShut
  {NoStop}%
\bibitem [{\citenamefont {Berezinsky}(1971)}]{Berezinsky71}%
  \BibitemOpen
  \bibfield  {author} {\bibinfo {author} {\bibfnamefont {V.~L.}\ \bibnamefont
  {Berezinsky}},\ }\bibfield  {title} {\bibinfo {title} {{Destruction of long
  range order in one-dimensional and two-dimensional systems having a
  continuous symmetry group. I. Classical systems}},\ }\href@noop {} {\bibfield
   {journal} {\bibinfo  {journal} {Sov. Phys. JETP}\ }\textbf {\bibinfo
  {volume} {32}},\ \bibinfo {pages} {493} (\bibinfo {year} {1971})},\ \bibinfo
  {note} {[Zh.\ Éksp.\ Teor.\ Fiz.\ \textbf{59}, 907 (1970)]}\BibitemShut
  {NoStop}%
\bibitem [{\citenamefont {Kosterlitz}\ and\ \citenamefont
  {Thouless}(1973)}]{Kosterlitz73}%
  \BibitemOpen
  \bibfield  {author} {\bibinfo {author} {\bibfnamefont {J.~M.}\ \bibnamefont
  {Kosterlitz}}\ and\ \bibinfo {author} {\bibfnamefont {D.~J.}\ \bibnamefont
  {Thouless}},\ }\bibfield  {title} {\bibinfo {title} {Ordering, metastability
  and phase transitions in two-dimensional systems},\ }\href
  {https://doi.org/10.1088/0022-3719/6/7/010} {\bibfield  {journal} {\bibinfo
  {journal} {Journal of Physics C: Solid State Physics}\ }\textbf {\bibinfo
  {volume} {6}},\ \bibinfo {pages} {1181} (\bibinfo {year} {1973})}\BibitemShut
  {NoStop}%
\bibitem [{\citenamefont {Peierls}(1936)}]{Peierls36}%
  \BibitemOpen
  \bibfield  {author} {\bibinfo {author} {\bibfnamefont {R.}~\bibnamefont
  {Peierls}},\ }\bibfield  {title} {\bibinfo {title} {On {Ising’s} model of
  ferromagnetism},\ }\href {https://doi.org/10.1017/S0305004100019174}
  {\bibfield  {journal} {\bibinfo  {journal} {Mathematical Proceedings of the
  Cambridge Philosophical Society}\ }\textbf {\bibinfo {volume} {32}},\
  \bibinfo {pages} {477–481} (\bibinfo {year} {1936})}\BibitemShut {NoStop}%
\bibitem [{\citenamefont {Onsager}(1944)}]{Onsager44}%
  \BibitemOpen
  \bibfield  {author} {\bibinfo {author} {\bibfnamefont {L.}~\bibnamefont
  {Onsager}},\ }\bibfield  {title} {\bibinfo {title} {Crystal statistics. i. a
  two-dimensional model with an order-disorder transition},\ }\href
  {https://doi.org/10.1103/PhysRev.65.117} {\bibfield  {journal} {\bibinfo
  {journal} {Phys. Rev.}\ }\textbf {\bibinfo {volume} {65}},\ \bibinfo {pages}
  {117} (\bibinfo {year} {1944})}\BibitemShut {NoStop}%
\bibitem [{\citenamefont {Matsubara}\ and\ \citenamefont
  {Matsuda}(1956)}]{Matsubara56}%
  \BibitemOpen
  \bibfield  {author} {\bibinfo {author} {\bibfnamefont {T.}~\bibnamefont
  {Matsubara}}\ and\ \bibinfo {author} {\bibfnamefont {H.}~\bibnamefont
  {Matsuda}},\ }\bibfield  {title} {\bibinfo {title} {{A Lattice Model of
  Liquid Helium}},\ }\href {https://doi.org/10.1143/PTP.16.416} {\bibfield
  {journal} {\bibinfo  {journal} {Progress of Theoretical Physics}\ }\textbf
  {\bibinfo {volume} {16}},\ \bibinfo {pages} {416} (\bibinfo {year} {1956})},\
  \Eprint
  {https://arxiv.org/abs/https://academic.oup.com/ptp/article-pdf/16/4/416/5226644/16-4-416.pdf}
  {https://academic.oup.com/ptp/article-pdf/16/4/416/5226644/16-4-416.pdf}
  \BibitemShut {NoStop}%
\bibitem [{\citenamefont {Suzuki}\ \emph {et~al.}(1977)\citenamefont {Suzuki},
  \citenamefont {Miyashita}, \citenamefont {Kuroda},\ and\ \citenamefont
  {Kawabata}}]{Suzuki77}%
  \BibitemOpen
  \bibfield  {author} {\bibinfo {author} {\bibfnamefont {M.}~\bibnamefont
  {Suzuki}}, \bibinfo {author} {\bibfnamefont {S.}~\bibnamefont {Miyashita}},
  \bibinfo {author} {\bibfnamefont {A.}~\bibnamefont {Kuroda}},\ and\ \bibinfo
  {author} {\bibfnamefont {C.}~\bibnamefont {Kawabata}},\ }\bibfield  {title}
  {\bibinfo {title} {Monte carlo simulations of the two-dimensional quantal and
  classical spin systems — a new type of phase transition with vortices},\
  }\href {https://doi.org/https://doi.org/10.1016/0375-9601(77)90061-5}
  {\bibfield  {journal} {\bibinfo  {journal} {Physics Letters A}\ }\textbf
  {\bibinfo {volume} {60}},\ \bibinfo {pages} {478} (\bibinfo {year}
  {1977})}\BibitemShut {NoStop}%
\bibitem [{\citenamefont {Pearson}(1977)}]{Pearson77}%
  \BibitemOpen
  \bibfield  {author} {\bibinfo {author} {\bibfnamefont {R.~B.}\ \bibnamefont
  {Pearson}},\ }\bibfield  {title} {\bibinfo {title} {Estimates of the
  ground-state eigenvalue of the two-dimensional spin-$\frac{1}{2}$
  $x\ensuremath{-}y$ model},\ }\href {https://doi.org/10.1103/PhysRevB.16.1109}
  {\bibfield  {journal} {\bibinfo  {journal} {Phys. Rev. B}\ }\textbf {\bibinfo
  {volume} {16}},\ \bibinfo {pages} {1109} (\bibinfo {year}
  {1977})}\BibitemShut {NoStop}%
\bibitem [{\citenamefont {Betts}\ and\ \citenamefont {Oitmaa}(1977)}]{Betts77}%
  \BibitemOpen
  \bibfield  {author} {\bibinfo {author} {\bibfnamefont {D.}~\bibnamefont
  {Betts}}\ and\ \bibinfo {author} {\bibfnamefont {J.}~\bibnamefont {Oitmaa}},\
  }\bibfield  {title} {\bibinfo {title} {On the ground state of the xy magnet
  and heisenberg antiferromagnet on the square lattice},\ }\href
  {https://doi.org/https://doi.org/10.1016/0375-9601(77)90796-4} {\bibfield
  {journal} {\bibinfo  {journal} {Physics Letters A}\ }\textbf {\bibinfo
  {volume} {62}},\ \bibinfo {pages} {277} (\bibinfo {year} {1977})}\BibitemShut
  {NoStop}%
\bibitem [{\citenamefont {Rogiers}\ \emph
  {et~al.}(1978{\natexlab{a}})\citenamefont {Rogiers}, \citenamefont {Lookman},
  \citenamefont {Betts},\ and\ \citenamefont {Elliott}}]{Rogiers78a}%
  \BibitemOpen
  \bibfield  {author} {\bibinfo {author} {\bibfnamefont {J.}~\bibnamefont
  {Rogiers}}, \bibinfo {author} {\bibfnamefont {T.}~\bibnamefont {Lookman}},
  \bibinfo {author} {\bibfnamefont {D.~D.}\ \bibnamefont {Betts}},\ and\
  \bibinfo {author} {\bibfnamefont {C.~J.}\ \bibnamefont {Elliott}},\
  }\bibfield  {title} {\bibinfo {title} {The spin model. i. derivation of high
  temperature series expansions for thermodynamic quantities},\ }\href
  {https://doi.org/10.1139/p78-053} {\bibfield  {journal} {\bibinfo  {journal}
  {Canadian Journal of Physics}\ }\textbf {\bibinfo {volume} {56}},\ \bibinfo
  {pages} {409} (\bibinfo {year} {1978}{\natexlab{a}})},\ \Eprint
  {https://arxiv.org/abs/https://doi.org/10.1139/p78-053}
  {https://doi.org/10.1139/p78-053} \BibitemShut {NoStop}%
\bibitem [{\citenamefont {Rogiers}\ \emph
  {et~al.}(1978{\natexlab{b}})\citenamefont {Rogiers}, \citenamefont {Betts},\
  and\ \citenamefont {Lookman}}]{Rogiers78b}%
  \BibitemOpen
  \bibfield  {author} {\bibinfo {author} {\bibfnamefont {J.}~\bibnamefont
  {Rogiers}}, \bibinfo {author} {\bibfnamefont {D.~D.}\ \bibnamefont {Betts}},\
  and\ \bibinfo {author} {\bibfnamefont {T.}~\bibnamefont {Lookman}},\
  }\bibfield  {title} {\bibinfo {title} {The spin model. ii. analysis of high
  temperature series expansions of some thermodynamic quantities in three
  dimensions},\ }\href {https://doi.org/10.1139/p78-054} {\bibfield  {journal}
  {\bibinfo  {journal} {Canadian Journal of Physics}\ }\textbf {\bibinfo
  {volume} {56}},\ \bibinfo {pages} {420} (\bibinfo {year}
  {1978}{\natexlab{b}})},\ \Eprint
  {https://arxiv.org/abs/https://doi.org/10.1139/p78-054}
  {https://doi.org/10.1139/p78-054} \BibitemShut {NoStop}%
\bibitem [{\citenamefont {Oitmaa}\ and\ \citenamefont
  {Betts}(1978{\natexlab{a}})}]{Oitmaa78a}%
  \BibitemOpen
  \bibfield  {author} {\bibinfo {author} {\bibfnamefont {J.}~\bibnamefont
  {Oitmaa}}\ and\ \bibinfo {author} {\bibfnamefont {D.~D.}\ \bibnamefont
  {Betts}},\ }\bibfield  {title} {\bibinfo {title} {The ground state of two
  quantum models of magnetism},\ }\href {https://doi.org/10.1139/p78-120}
  {\bibfield  {journal} {\bibinfo  {journal} {Canadian Journal of Physics}\
  }\textbf {\bibinfo {volume} {56}},\ \bibinfo {pages} {897} (\bibinfo {year}
  {1978}{\natexlab{a}})},\ \Eprint
  {https://arxiv.org/abs/https://doi.org/10.1139/p78-120}
  {https://doi.org/10.1139/p78-120} \BibitemShut {NoStop}%
\bibitem [{\citenamefont {Oitmaa}\ and\ \citenamefont
  {Betts}(1978{\natexlab{b}})}]{Oitmaa78b}%
  \BibitemOpen
  \bibfield  {author} {\bibinfo {author} {\bibfnamefont {J.}~\bibnamefont
  {Oitmaa}}\ and\ \bibinfo {author} {\bibfnamefont {D.}~\bibnamefont {Betts}},\
  }\bibfield  {title} {\bibinfo {title} {Ground state properties of the s = 12
  heisenberg antiferromagnet and xy magnet in three dimensions},\ }\href
  {https://doi.org/https://doi.org/10.1016/0375-9601(78)90625-4} {\bibfield
  {journal} {\bibinfo  {journal} {Physics Letters A}\ }\textbf {\bibinfo
  {volume} {68}},\ \bibinfo {pages} {450} (\bibinfo {year}
  {1978}{\natexlab{b}})}\BibitemShut {NoStop}%
\bibitem [{\citenamefont {Suzuki}\ and\ \citenamefont
  {Miyashita}(1978)}]{Suzuki78}%
  \BibitemOpen
  \bibfield  {author} {\bibinfo {author} {\bibfnamefont {M.}~\bibnamefont
  {Suzuki}}\ and\ \bibinfo {author} {\bibfnamefont {S.}~\bibnamefont
  {Miyashita}},\ }\bibfield  {title} {\bibinfo {title} {Variational study on
  the ground state of the spin xy magnet},\ }\href
  {https://doi.org/10.1139/p78-121} {\bibfield  {journal} {\bibinfo  {journal}
  {Canadian Journal of Physics}\ }\textbf {\bibinfo {volume} {56}},\ \bibinfo
  {pages} {902} (\bibinfo {year} {1978})},\ \Eprint
  {https://arxiv.org/abs/https://doi.org/10.1139/p78-121}
  {https://doi.org/10.1139/p78-121} \BibitemShut {NoStop}%
\bibitem [{\citenamefont {Uchinami}\ \emph {et~al.}(1979)\citenamefont
  {Uchinami}, \citenamefont {Takada},\ and\ \citenamefont
  {Takano}}]{Uchinami79}%
  \BibitemOpen
  \bibfield  {author} {\bibinfo {author} {\bibfnamefont {M.}~\bibnamefont
  {Uchinami}}, \bibinfo {author} {\bibfnamefont {S.}~\bibnamefont {Takada}},\
  and\ \bibinfo {author} {\bibfnamefont {F.}~\bibnamefont {Takano}},\
  }\bibfield  {title} {\bibinfo {title} {Spin wave theory of the spin 1/2 xy
  model},\ }\href {https://doi.org/10.1143/JPSJ.47.1047} {\bibfield  {journal}
  {\bibinfo  {journal} {Journal of the Physical Society of Japan}\ }\textbf
  {\bibinfo {volume} {47}},\ \bibinfo {pages} {1047} (\bibinfo {year}
  {1979})},\ \Eprint
  {https://arxiv.org/abs/https://doi.org/10.1143/JPSJ.47.1047}
  {https://doi.org/10.1143/JPSJ.47.1047} \BibitemShut {NoStop}%
\bibitem [{\citenamefont {Oitmaa}\ \emph {et~al.}(1980)\citenamefont {Oitmaa},
  \citenamefont {Betts},\ and\ \citenamefont {Marland}}]{Oitmaa80}%
  \BibitemOpen
  \bibfield  {author} {\bibinfo {author} {\bibfnamefont {J.}~\bibnamefont
  {Oitmaa}}, \bibinfo {author} {\bibfnamefont {D.}~\bibnamefont {Betts}},\ and\
  \bibinfo {author} {\bibfnamefont {L.}~\bibnamefont {Marland}},\ }\bibfield
  {title} {\bibinfo {title} {Susceptibilities of the s=12 xy model on the
  square lattice at t = 0},\ }\href
  {https://doi.org/https://doi.org/10.1016/0375-9601(80)90245-5} {\bibfield
  {journal} {\bibinfo  {journal} {Physics Letters A}\ }\textbf {\bibinfo
  {volume} {79}},\ \bibinfo {pages} {193} (\bibinfo {year} {1980})}\BibitemShut
  {NoStop}%
\bibitem [{\citenamefont {Deraedt}\ \emph {et~al.}(1984)\citenamefont
  {Deraedt}, \citenamefont {Deraedt},\ and\ \citenamefont
  {Lagendijk}}]{Deraedt84a}%
  \BibitemOpen
  \bibfield  {author} {\bibinfo {author} {\bibfnamefont {H.}~\bibnamefont
  {Deraedt}}, \bibinfo {author} {\bibfnamefont {B.}~\bibnamefont {Deraedt}},\
  and\ \bibinfo {author} {\bibfnamefont {A.}~\bibnamefont {Lagendijk}},\
  }\bibfield  {title} {\bibinfo {title} {Thermodynamics of the 2-dimensional
  spin-1/2 xy model},\ }\href
  {https://doi.org/https://doi.org/10.1007/BF01318413} {\bibfield  {journal}
  {\bibinfo  {journal} {Zeitschrift fur Physik B-Condensed Matter}\ }\textbf
  {\bibinfo {volume} {57}},\ \bibinfo {pages} {209} (\bibinfo {year}
  {1984})}\BibitemShut {NoStop}%
\bibitem [{\citenamefont {{De Raedt}}\ \emph {et~al.}(1984)\citenamefont {{De
  Raedt}}, \citenamefont {{De Raedt}}, \citenamefont {Fivez},\ and\
  \citenamefont {Lagendijk}}]{Deraedt84b}%
  \BibitemOpen
  \bibfield  {author} {\bibinfo {author} {\bibfnamefont {H.}~\bibnamefont {{De
  Raedt}}}, \bibinfo {author} {\bibfnamefont {B.}~\bibnamefont {{De Raedt}}},
  \bibinfo {author} {\bibfnamefont {J.}~\bibnamefont {Fivez}},\ and\ \bibinfo
  {author} {\bibfnamefont {A.}~\bibnamefont {Lagendijk}},\ }\bibfield  {title}
  {\bibinfo {title} {Monte carlo study of the two-dimensional spin-1/2 xy
  model},\ }\href
  {https://doi.org/https://doi.org/10.1016/0375-9601(84)90750-3} {\bibfield
  {journal} {\bibinfo  {journal} {Physics Letters A}\ }\textbf {\bibinfo
  {volume} {104}},\ \bibinfo {pages} {430} (\bibinfo {year}
  {1984})}\BibitemShut {NoStop}%
\bibitem [{\citenamefont {Loh}\ \emph {et~al.}(1985)\citenamefont {Loh},
  \citenamefont {Scalapino},\ and\ \citenamefont {Grant}}]{Loh85}%
  \BibitemOpen
  \bibfield  {author} {\bibinfo {author} {\bibfnamefont {E.}~\bibnamefont
  {Loh}}, \bibinfo {author} {\bibfnamefont {D.~J.}\ \bibnamefont {Scalapino}},\
  and\ \bibinfo {author} {\bibfnamefont {P.~M.}\ \bibnamefont {Grant}},\
  }\bibfield  {title} {\bibinfo {title} {Monte carlo studies of the quantum
  $\mathrm{XY}$ model in two dimensions},\ }\href
  {https://doi.org/10.1103/PhysRevB.31.4712} {\bibfield  {journal} {\bibinfo
  {journal} {Phys. Rev. B}\ }\textbf {\bibinfo {volume} {31}},\ \bibinfo
  {pages} {4712} (\bibinfo {year} {1985})}\BibitemShut {NoStop}%
\bibitem [{\citenamefont {Takahashi}(1986)}]{Takahashi86}%
  \BibitemOpen
  \bibfield  {author} {\bibinfo {author} {\bibfnamefont {M.}~\bibnamefont
  {Takahashi}},\ }\bibfield  {title} {\bibinfo {title} {{Quantum Heisenberg
  Ferromagnets in One and Two Dimensions at Low Temperature}},\ }\href
  {https://doi.org/10.1143/PTPS.87.233} {\bibfield  {journal} {\bibinfo
  {journal} {Progress of Theoretical Physics Supplement}\ }\textbf {\bibinfo
  {volume} {87}},\ \bibinfo {pages} {233} (\bibinfo {year} {1986})},\ \Eprint
  {https://arxiv.org/abs/https://academic.oup.com/ptps/article-pdf/doi/10.1143/PTPS.87.233/5297453/87-233.pdf}
  {https://academic.oup.com/ptps/article-pdf/doi/10.1143/PTPS.87.233/5297453/87-233.pdf}
  \BibitemShut {NoStop}%
\bibitem [{\citenamefont {Gomez-Santos}\ and\ \citenamefont
  {Joannopoulos}(1987)}]{Gomez-Santos87}%
  \BibitemOpen
  \bibfield  {author} {\bibinfo {author} {\bibfnamefont {G.}~\bibnamefont
  {Gomez-Santos}}\ and\ \bibinfo {author} {\bibfnamefont {J.~D.}\ \bibnamefont
  {Joannopoulos}},\ }\bibfield  {title} {\bibinfo {title} {Application of
  spin-wave theory to the ground state of xy quantum hamiltonians},\ }\href
  {https://doi.org/10.1103/PhysRevB.36.8707} {\bibfield  {journal} {\bibinfo
  {journal} {Phys. Rev. B}\ }\textbf {\bibinfo {volume} {36}},\ \bibinfo
  {pages} {8707} (\bibinfo {year} {1987})}\BibitemShut {NoStop}%
\bibitem [{\citenamefont {Tang}(1988)}]{Tang88}%
  \BibitemOpen
  \bibfield  {author} {\bibinfo {author} {\bibfnamefont {S.}~\bibnamefont
  {Tang}},\ }\bibfield  {title} {\bibinfo {title} {Zero-temperature ordering in
  the two-dimensional quantum xy model},\ }\href
  {https://doi.org/https://doi.org/10.1016/0375-9601(88)90074-6} {\bibfield
  {journal} {\bibinfo  {journal} {Physics Letters A}\ }\textbf {\bibinfo
  {volume} {129}},\ \bibinfo {pages} {83} (\bibinfo {year} {1988})}\BibitemShut
  {NoStop}%
\bibitem [{\citenamefont {Okabe}\ and\ \citenamefont
  {Kikuchi}(1988)}]{Okabe88}%
  \BibitemOpen
  \bibfield  {author} {\bibinfo {author} {\bibfnamefont {Y.}~\bibnamefont
  {Okabe}}\ and\ \bibinfo {author} {\bibfnamefont {M.}~\bibnamefont
  {Kikuchi}},\ }\bibfield  {title} {\bibinfo {title} {Quantum monte carlo
  simulation of the spin 1/2 xxz model on the square lattice},\ }\href
  {https://doi.org/10.1143/JPSJ.57.4351} {\bibfield  {journal} {\bibinfo
  {journal} {Journal of the Physical Society of Japan}\ }\textbf {\bibinfo
  {volume} {57}},\ \bibinfo {pages} {4351} (\bibinfo {year} {1988})},\ \Eprint
  {https://arxiv.org/abs/https://doi.org/10.1143/JPSJ.57.4351}
  {https://doi.org/10.1143/JPSJ.57.4351} \BibitemShut {NoStop}%
\bibitem [{\citenamefont {Kennedy}\ \emph {et~al.}(1988)\citenamefont
  {Kennedy}, \citenamefont {Lieb},\ and\ \citenamefont {Shastry}}]{Kennedy88}%
  \BibitemOpen
  \bibfield  {author} {\bibinfo {author} {\bibfnamefont {T.}~\bibnamefont
  {Kennedy}}, \bibinfo {author} {\bibfnamefont {E.~H.}\ \bibnamefont {Lieb}},\
  and\ \bibinfo {author} {\bibfnamefont {B.~S.}\ \bibnamefont {Shastry}},\
  }\bibfield  {title} {\bibinfo {title} {The $\mathrm{XY}$ model has long-range
  order for all spins and all dimensions greater than one},\ }\href
  {https://doi.org/10.1103/PhysRevLett.61.2582} {\bibfield  {journal} {\bibinfo
   {journal} {Phys. Rev. Lett.}\ }\textbf {\bibinfo {volume} {61}},\ \bibinfo
  {pages} {2582} (\bibinfo {year} {1988})}\BibitemShut {NoStop}%
\bibitem [{\citenamefont {Chakravarty}\ \emph {et~al.}(1989)\citenamefont
  {Chakravarty}, \citenamefont {Halperin},\ and\ \citenamefont
  {Nelson}}]{Chakravarty89}%
  \BibitemOpen
  \bibfield  {author} {\bibinfo {author} {\bibfnamefont {S.}~\bibnamefont
  {Chakravarty}}, \bibinfo {author} {\bibfnamefont {B.~I.}\ \bibnamefont
  {Halperin}},\ and\ \bibinfo {author} {\bibfnamefont {D.~R.}\ \bibnamefont
  {Nelson}},\ }\bibfield  {title} {\bibinfo {title} {Two-dimensional quantum
  heisenberg antiferromagnet at low temperatures},\ }\href
  {https://doi.org/10.1103/PhysRevB.39.2344} {\bibfield  {journal} {\bibinfo
  {journal} {Phys. Rev. B}\ }\textbf {\bibinfo {volume} {39}},\ \bibinfo
  {pages} {2344} (\bibinfo {year} {1989})}\BibitemShut {NoStop}%
\bibitem [{\citenamefont {Drzewinski}\ and\ \citenamefont
  {Sznajd}(1989)}]{Drzewinski89}%
  \BibitemOpen
  \bibfield  {author} {\bibinfo {author} {\bibfnamefont {A.}~\bibnamefont
  {Drzewinski}}\ and\ \bibinfo {author} {\bibfnamefont {J.}~\bibnamefont
  {Sznajd}},\ }\bibfield  {title} {\bibinfo {title} {Real space renormalization
  group study of the anisotropic quantum heisenberg model on a square
  lattice},\ }\href
  {https://doi.org/https://doi.org/10.1016/0375-9601(89)90880-3} {\bibfield
  {journal} {\bibinfo  {journal} {Physics Letters A}\ }\textbf {\bibinfo
  {volume} {138}},\ \bibinfo {pages} {143} (\bibinfo {year}
  {1989})}\BibitemShut {NoStop}%
\bibitem [{\citenamefont {Ding}\ and\ \citenamefont
  {Makivi\ifmmode~\acute{c}\else \'{c}\fi{}}(1990)}]{Ding90a}%
  \BibitemOpen
  \bibfield  {author} {\bibinfo {author} {\bibfnamefont {H.-Q.}\ \bibnamefont
  {Ding}}\ and\ \bibinfo {author} {\bibfnamefont {M.~S.}\ \bibnamefont
  {Makivi\ifmmode~\acute{c}\else \'{c}\fi{}}},\ }\bibfield  {title} {\bibinfo
  {title} {Kosterlitz-thouless transition in the two-dimensional quantum xy
  model},\ }\href {https://doi.org/10.1103/PhysRevB.42.6827} {\bibfield
  {journal} {\bibinfo  {journal} {Phys. Rev. B}\ }\textbf {\bibinfo {volume}
  {42}},\ \bibinfo {pages} {6827} (\bibinfo {year} {1990})}\BibitemShut
  {NoStop}%
\bibitem [{\citenamefont {Hamer}\ \emph {et~al.}(1991)\citenamefont {Hamer},
  \citenamefont {Oitmaa},\ and\ \citenamefont {Weihong}}]{Hamer91}%
  \BibitemOpen
  \bibfield  {author} {\bibinfo {author} {\bibfnamefont {C.~J.}\ \bibnamefont
  {Hamer}}, \bibinfo {author} {\bibfnamefont {J.}~\bibnamefont {Oitmaa}},\ and\
  \bibinfo {author} {\bibfnamefont {Z.}~\bibnamefont {Weihong}},\ }\bibfield
  {title} {\bibinfo {title} {Zero-temperature properties of the quantum xy
  model with anisotropy},\ }\href {https://doi.org/10.1103/PhysRevB.43.10789}
  {\bibfield  {journal} {\bibinfo  {journal} {Phys. Rev. B}\ }\textbf {\bibinfo
  {volume} {43}},\ \bibinfo {pages} {10789} (\bibinfo {year}
  {1991})}\BibitemShut {NoStop}%
\bibitem [{\citenamefont {Zhang}\ and\ \citenamefont {Runge}(1992)}]{Zhang92}%
  \BibitemOpen
  \bibfield  {author} {\bibinfo {author} {\bibfnamefont {S.}~\bibnamefont
  {Zhang}}\ and\ \bibinfo {author} {\bibfnamefont {K.~J.}\ \bibnamefont
  {Runge}},\ }\bibfield  {title} {\bibinfo {title} {Green's-function monte
  carlo study of the two-dimensional, spin-1/2 xy ferromagnet},\ }\href
  {https://doi.org/10.1103/PhysRevB.45.1052} {\bibfield  {journal} {\bibinfo
  {journal} {Phys. Rev. B}\ }\textbf {\bibinfo {volume} {45}},\ \bibinfo
  {pages} {1052} (\bibinfo {year} {1992})}\BibitemShut {NoStop}%
\bibitem [{\citenamefont {Ding}(1992{\natexlab{a}})}]{Ding92a}%
  \BibitemOpen
  \bibfield  {author} {\bibinfo {author} {\bibfnamefont {H.-Q.}\ \bibnamefont
  {Ding}},\ }\bibfield  {title} {\bibinfo {title} {Phase transition and
  thermodynamics of quantum xy model in two dimensions},\ }\href
  {https://doi.org/10.1103/PhysRevB.45.230} {\bibfield  {journal} {\bibinfo
  {journal} {Phys. Rev. B}\ }\textbf {\bibinfo {volume} {45}},\ \bibinfo
  {pages} {230} (\bibinfo {year} {1992}{\natexlab{a}})}\BibitemShut {NoStop}%
\bibitem [{\citenamefont
  {Makivic\ifmmode\acute\else\textasciiacute\fi{}}(1992)}]{Makivic92}%
  \BibitemOpen
  \bibfield  {author} {\bibinfo {author} {\bibfnamefont {M.~S.}\ \bibnamefont
  {Makivic\ifmmode\acute\else\textasciiacute\fi{}}},\ }\bibfield  {title}
  {\bibinfo {title} {Low temperature phase of the two-dimensional quantum xy
  model},\ }\href {https://doi.org/10.1103/PhysRevB.46.3167} {\bibfield
  {journal} {\bibinfo  {journal} {Phys. Rev. B}\ }\textbf {\bibinfo {volume}
  {46}},\ \bibinfo {pages} {3167} (\bibinfo {year} {1992})}\BibitemShut
  {NoStop}%
\bibitem [{\citenamefont {Hasenfratz}\ and\ \citenamefont
  {Niedermayer}(1993)}]{Hasenfratz93}%
  \BibitemOpen
  \bibfield  {author} {\bibinfo {author} {\bibfnamefont {P.}~\bibnamefont
  {Hasenfratz}}\ and\ \bibinfo {author} {\bibfnamefont {F.}~\bibnamefont
  {Niedermayer}},\ }\bibfield  {title} {\bibinfo {title} {Finite-size and
  temperature effects in the af heisenberg-model},\ }\href
  {https://doi.org/10.1007/BF01309171} {\bibfield  {journal} {\bibinfo
  {journal} {Zeitschrift fur Physik B-Condensed Matter}\ }\textbf {\bibinfo
  {volume} {92}},\ \bibinfo {pages} {91} (\bibinfo {year} {1993})}\BibitemShut
  {NoStop}%
\bibitem [{\citenamefont {Daren}\ and\ \citenamefont {Heping}(1994)}]{Ji94}%
  \BibitemOpen
  \bibfield  {author} {\bibinfo {author} {\bibfnamefont {J.}~\bibnamefont
  {Daren}}\ and\ \bibinfo {author} {\bibfnamefont {Y.}~\bibnamefont {Heping}},\
  }\bibfield  {title} {\bibinfo {title} {Monte carlo simulations of the quantum
  x-y model by a loop-cluster algorithm},\ }\href
  {https://doi.org/10.1088/0256-307X/11/1/014} {\bibfield  {journal} {\bibinfo
  {journal} {Chinese Physics Letters}\ }\textbf {\bibinfo {volume} {11}},\
  \bibinfo {pages} {49} (\bibinfo {year} {1994})}\BibitemShut {NoStop}%
\bibitem [{\citenamefont {Pires}(1996)}]{Pires96}%
  \BibitemOpen
  \bibfield  {author} {\bibinfo {author} {\bibfnamefont {A.~S.~T.}\
  \bibnamefont {Pires}},\ }\bibfield  {title} {\bibinfo {title}
  {Low-temperature thermodynamic study of the xy model using a self-consistent
  harmonic approximation},\ }\href {https://doi.org/10.1103/PhysRevB.53.235}
  {\bibfield  {journal} {\bibinfo  {journal} {Phys. Rev. B}\ }\textbf {\bibinfo
  {volume} {53}},\ \bibinfo {pages} {235} (\bibinfo {year} {1996})}\BibitemShut
  {NoStop}%
\bibitem [{\citenamefont {Harada}\ and\ \citenamefont
  {Kawashima}(1998)}]{Harada98}%
  \BibitemOpen
  \bibfield  {author} {\bibinfo {author} {\bibfnamefont {K.}~\bibnamefont
  {Harada}}\ and\ \bibinfo {author} {\bibfnamefont {N.}~\bibnamefont
  {Kawashima}},\ }\bibfield  {title} {\bibinfo {title} {Kosterlitz-thouless
  transition of quantum xy model in two dimensions},\ }\href
  {https://doi.org/10.1143/JPSJ.67.2768} {\bibfield  {journal} {\bibinfo
  {journal} {Journal of the Physical Society of Japan}\ }\textbf {\bibinfo
  {volume} {67}},\ \bibinfo {pages} {2768} (\bibinfo {year} {1998})},\ \Eprint
  {https://arxiv.org/abs/https://doi.org/10.1143/JPSJ.67.2768}
  {https://doi.org/10.1143/JPSJ.67.2768} \BibitemShut {NoStop}%
\bibitem [{\citenamefont {Ying}\ \emph {et~al.}(1998)\citenamefont {Ying},
  \citenamefont {Luo}, \citenamefont {Sch\"{u}lke},\ and\ \citenamefont
  {Zheng}}]{Ying98}%
  \BibitemOpen
  \bibfield  {author} {\bibinfo {author} {\bibfnamefont {H.~P.}\ \bibnamefont
  {Ying}}, \bibinfo {author} {\bibfnamefont {H.~J.}\ \bibnamefont {Luo}},
  \bibinfo {author} {\bibfnamefont {L.}~\bibnamefont {Sch\"{u}lke}},\ and\
  \bibinfo {author} {\bibfnamefont {B.}~\bibnamefont {Zheng}},\ }\bibfield
  {title} {\bibinfo {title} {Dynamic monte carlo study of the two-dimensional
  quantum xy model},\ }\href {https://doi.org/10.1142/S0217984998001463}
  {\bibfield  {journal} {\bibinfo  {journal} {Modern Physics Letters B}\
  }\textbf {\bibinfo {volume} {12}},\ \bibinfo {pages} {1237} (\bibinfo {year}
  {1998})},\ \Eprint
  {https://arxiv.org/abs/https://doi.org/10.1142/S0217984998001463}
  {https://doi.org/10.1142/S0217984998001463} \BibitemShut {NoStop}%
\bibitem [{\citenamefont {Hamer}\ \emph {et~al.}(1999)\citenamefont {Hamer},
  \citenamefont {Hövelborn},\ and\ \citenamefont {Bachhuber}}]{Hamer99}%
  \BibitemOpen
  \bibfield  {author} {\bibinfo {author} {\bibfnamefont {C.~J.}\ \bibnamefont
  {Hamer}}, \bibinfo {author} {\bibfnamefont {T.}~\bibnamefont {Hövelborn}},\
  and\ \bibinfo {author} {\bibfnamefont {M.}~\bibnamefont {Bachhuber}},\
  }\bibfield  {title} {\bibinfo {title} {Finite-size scaling in the spin- model
  on a square lattice},\ }\href {https://doi.org/10.1088/0305-4470/32/1/007}
  {\bibfield  {journal} {\bibinfo  {journal} {Journal of Physics A:
  Mathematical and General}\ }\textbf {\bibinfo {volume} {32}},\ \bibinfo
  {pages} {51} (\bibinfo {year} {1999})}\BibitemShut {NoStop}%
\bibitem [{\citenamefont {Sandvik}\ and\ \citenamefont
  {Hamer}(1999)}]{Sandvik99}%
  \BibitemOpen
  \bibfield  {author} {\bibinfo {author} {\bibfnamefont {A.~W.}\ \bibnamefont
  {Sandvik}}\ and\ \bibinfo {author} {\bibfnamefont {C.~J.}\ \bibnamefont
  {Hamer}},\ }\bibfield  {title} {\bibinfo {title} {Ground-state parameters,
  finite-size scaling, and low-temperature properties of the two-dimensional
  $s=\frac{1}{2}$ $\mathrm{XY}$ model},\ }\href
  {https://doi.org/10.1103/PhysRevB.60.6588} {\bibfield  {journal} {\bibinfo
  {journal} {Phys. Rev. B}\ }\textbf {\bibinfo {volume} {60}},\ \bibinfo
  {pages} {6588} (\bibinfo {year} {1999})}\BibitemShut {NoStop}%
\bibitem [{\citenamefont {Schindelin}\ \emph {et~al.}(2001)\citenamefont
  {Schindelin}, \citenamefont {Fehske}, \citenamefont {Büttner},\ and\
  \citenamefont {Ihle}}]{Schindelin01}%
  \BibitemOpen
  \bibfield  {author} {\bibinfo {author} {\bibfnamefont {C.}~\bibnamefont
  {Schindelin}}, \bibinfo {author} {\bibfnamefont {H.}~\bibnamefont {Fehske}},
  \bibinfo {author} {\bibfnamefont {H.}~\bibnamefont {Büttner}},\ and\
  \bibinfo {author} {\bibfnamefont {D.}~\bibnamefont {Ihle}},\ }\bibfield
  {title} {\bibinfo {title} {Quantum effects in the 2d xy model},\ }\href
  {https://doi.org/https://doi.org/10.1016/S0304-8853(00)01193-8} {\bibfield
  {journal} {\bibinfo  {journal} {Journal of Magnetism and Magnetic Materials}\
  }\textbf {\bibinfo {volume} {226-230}},\ \bibinfo {pages} {403} (\bibinfo
  {year} {2001})},\ \bibinfo {note} {proceedings of the International
  Conference on Magnetism (ICM 2000)}\BibitemShut {NoStop}%
\bibitem [{\citenamefont {Melko}\ \emph {et~al.}(2004)\citenamefont {Melko},
  \citenamefont {Sandvik},\ and\ \citenamefont {Scalapino}}]{Melko04}%
  \BibitemOpen
  \bibfield  {author} {\bibinfo {author} {\bibfnamefont {R.~G.}\ \bibnamefont
  {Melko}}, \bibinfo {author} {\bibfnamefont {A.~W.}\ \bibnamefont {Sandvik}},\
  and\ \bibinfo {author} {\bibfnamefont {D.~J.}\ \bibnamefont {Scalapino}},\
  }\bibfield  {title} {\bibinfo {title} {Aspect-ratio dependence of the spin
  stiffness of a two-dimensional $\mathrm{XY}$ model},\ }\href
  {https://doi.org/10.1103/PhysRevB.69.014509} {\bibfield  {journal} {\bibinfo
  {journal} {Phys. Rev. B}\ }\textbf {\bibinfo {volume} {69}},\ \bibinfo
  {pages} {014509} (\bibinfo {year} {2004})}\BibitemShut {NoStop}%
\bibitem [{\citenamefont {Carrasquilla}\ and\ \citenamefont
  {Rigol}(2012)}]{Carrasquilla12}%
  \BibitemOpen
  \bibfield  {author} {\bibinfo {author} {\bibfnamefont {J.}~\bibnamefont
  {Carrasquilla}}\ and\ \bibinfo {author} {\bibfnamefont {M.}~\bibnamefont
  {Rigol}},\ }\bibfield  {title} {\bibinfo {title} {Superfluid to normal phase
  transition in strongly correlated bosons in two and three dimensions},\
  }\href {https://doi.org/10.1103/PhysRevA.86.043629} {\bibfield  {journal}
  {\bibinfo  {journal} {Phys. Rev. A}\ }\textbf {\bibinfo {volume} {86}},\
  \bibinfo {pages} {043629} (\bibinfo {year} {2012})}\BibitemShut {NoStop}%
\bibitem [{\citenamefont {Hofmann}(2014)}]{Hofmann14}%
  \BibitemOpen
  \bibfield  {author} {\bibinfo {author} {\bibfnamefont {C.~P.}\ \bibnamefont
  {Hofmann}},\ }\bibfield  {title} {\bibinfo {title} {Systematic effective
  field theory analysis of the d=2+1 quantum xy model at low temperatures},\
  }\href {https://doi.org/10.1088/1742-5468/2014/02/P02006} {\bibfield
  {journal} {\bibinfo  {journal} {Journal of Statistical Mechanics: Theory and
  Experiment}\ }\textbf {\bibinfo {volume} {2014}},\ \bibinfo {pages} {P02006}
  (\bibinfo {year} {2014})}\BibitemShut {NoStop}%
\bibitem [{\citenamefont {Hofmann}(2016)}]{Hofmann16}%
  \BibitemOpen
  \bibfield  {author} {\bibinfo {author} {\bibfnamefont {C.~P.}\ \bibnamefont
  {Hofmann}},\ }\bibfield  {title} {\bibinfo {title} {(pseudo-)goldstone boson
  interaction in d=2+1 systems with a spontaneously broken internal rotation
  symmetry},\ }\href
  {https://doi.org/https://doi.org/10.1016/j.nuclphysb.2016.01.018} {\bibfield
  {journal} {\bibinfo  {journal} {Nuclear Physics B}\ }\textbf {\bibinfo
  {volume} {904}},\ \bibinfo {pages} {348} (\bibinfo {year}
  {2016})}\BibitemShut {NoStop}%
\bibitem [{\citenamefont {Bishop}\ \emph {et~al.}(2017)\citenamefont {Bishop},
  \citenamefont {Li}, \citenamefont {Zinke}, \citenamefont {Darradi},
  \citenamefont {Richter}, \citenamefont {Farnell},\ and\ \citenamefont
  {Schulenburg}}]{Bishop17}%
  \BibitemOpen
  \bibfield  {author} {\bibinfo {author} {\bibfnamefont {R.}~\bibnamefont
  {Bishop}}, \bibinfo {author} {\bibfnamefont {P.}~\bibnamefont {Li}}, \bibinfo
  {author} {\bibfnamefont {R.}~\bibnamefont {Zinke}}, \bibinfo {author}
  {\bibfnamefont {R.}~\bibnamefont {Darradi}}, \bibinfo {author} {\bibfnamefont
  {J.}~\bibnamefont {Richter}}, \bibinfo {author} {\bibfnamefont
  {D.}~\bibnamefont {Farnell}},\ and\ \bibinfo {author} {\bibfnamefont
  {J.}~\bibnamefont {Schulenburg}},\ }\bibfield  {title} {\bibinfo {title} {The
  spin-half xxz antiferromagnet on the square lattice revisited: A high-order
  coupled cluster treatment},\ }\href
  {https://doi.org/https://doi.org/10.1016/j.jmmm.2016.11.043} {\bibfield
  {journal} {\bibinfo  {journal} {Journal of Magnetism and Magnetic Materials}\
  }\textbf {\bibinfo {volume} {428}},\ \bibinfo {pages} {178} (\bibinfo {year}
  {2017})}\BibitemShut {NoStop}%
\bibitem [{\citenamefont {Nambu}(1960)}]{Nambu60}%
  \BibitemOpen
  \bibfield  {author} {\bibinfo {author} {\bibfnamefont {Y.}~\bibnamefont
  {Nambu}},\ }\bibfield  {title} {\bibinfo {title} {Quasi-particles and gauge
  invariance in the theory of superconductivity},\ }\href
  {https://doi.org/10.1103/PhysRev.117.648} {\bibfield  {journal} {\bibinfo
  {journal} {Phys. Rev.}\ }\textbf {\bibinfo {volume} {117}},\ \bibinfo {pages}
  {648} (\bibinfo {year} {1960})}\BibitemShut {NoStop}%
\bibitem [{\citenamefont {Goldstone}(1961)}]{Goldstone61}%
  \BibitemOpen
  \bibfield  {author} {\bibinfo {author} {\bibfnamefont {J.}~\bibnamefont
  {Goldstone}},\ }\bibfield  {title} {\bibinfo {title} {{Field Theories with
  Superconductor Solutions}},\ }\href {https://doi.org/10.1007/BF02812722}
  {\bibfield  {journal} {\bibinfo  {journal} {Nuovo Cim.}\ }\textbf {\bibinfo
  {volume} {19}},\ \bibinfo {pages} {154} (\bibinfo {year} {1961})}\BibitemShut
  {NoStop}%
\bibitem [{\citenamefont {Usman}\ \emph {et~al.}(2015)\citenamefont {Usman},
  \citenamefont {Ilyas},\ and\ \citenamefont {Khan}}]{Usman15}%
  \BibitemOpen
  \bibfield  {author} {\bibinfo {author} {\bibfnamefont {M.}~\bibnamefont
  {Usman}}, \bibinfo {author} {\bibfnamefont {A.}~\bibnamefont {Ilyas}},\ and\
  \bibinfo {author} {\bibfnamefont {K.}~\bibnamefont {Khan}},\ }\bibfield
  {title} {\bibinfo {title} {Quantum renormalization group of the $\mathit{XY}$
  model in two dimensions},\ }\href
  {https://doi.org/10.1103/PhysRevA.92.032327} {\bibfield  {journal} {\bibinfo
  {journal} {Phys. Rev. A}\ }\textbf {\bibinfo {volume} {92}},\ \bibinfo
  {pages} {032327} (\bibinfo {year} {2015})}\BibitemShut {NoStop}%
\bibitem [{\citenamefont {Wu}\ and\ \citenamefont {Xu}(2016)}]{Wu16}%
  \BibitemOpen
  \bibfield  {author} {\bibinfo {author} {\bibfnamefont {W.}~\bibnamefont
  {Wu}}\ and\ \bibinfo {author} {\bibfnamefont {J.-B.}\ \bibnamefont {Xu}},\
  }\bibfield  {title} {\bibinfo {title} {Renormalization of trace distance and
  multipartite entanglement close to the quantum phase transitions of one- and
  two-dimensional spin-chain systems},\ }\href
  {https://doi.org/10.1209/0295-5075/115/40006} {\bibfield  {journal} {\bibinfo
   {journal} {Europhysics Letters}\ }\textbf {\bibinfo {volume} {115}},\
  \bibinfo {pages} {40006} (\bibinfo {year} {2016})}\BibitemShut {NoStop}%
\bibitem [{\citenamefont {Farajollahpour}\ and\ \citenamefont
  {Jafari}(2018)}]{Farajollahpour18}%
  \BibitemOpen
  \bibfield  {author} {\bibinfo {author} {\bibfnamefont {T.}~\bibnamefont
  {Farajollahpour}}\ and\ \bibinfo {author} {\bibfnamefont {S.~A.}\
  \bibnamefont {Jafari}},\ }\bibfield  {title} {\bibinfo {title} {Topological
  phase transition of the anisotropic $xy$ model with dzyaloshinskii-moriya
  interaction},\ }\href {https://doi.org/10.1103/PhysRevB.98.085136} {\bibfield
   {journal} {\bibinfo  {journal} {Phys. Rev. B}\ }\textbf {\bibinfo {volume}
  {98}},\ \bibinfo {pages} {085136} (\bibinfo {year} {2018})}\BibitemShut
  {NoStop}%
\bibitem [{\citenamefont {Zhao}\ \emph {et~al.}(2021)\citenamefont {Zhao},
  \citenamefont {Gao}, \citenamefont {Ren},\ and\ \citenamefont
  {Liu}}]{Zhao21}%
  \BibitemOpen
  \bibfield  {author} {\bibinfo {author} {\bibfnamefont {Z.~J.}\ \bibnamefont
  {Zhao}}, \bibinfo {author} {\bibfnamefont {G.~J.}\ \bibnamefont {Gao}},
  \bibinfo {author} {\bibfnamefont {Y.~P.}\ \bibnamefont {Ren}},\ and\ \bibinfo
  {author} {\bibfnamefont {X.~M.}\ \bibnamefont {Liu}},\ }\bibfield  {title}
  {\bibinfo {title} {Renormalization-group approach to quantum fisher
  information in the two-dimensional xy model},\ }\href
  {https://doi.org/10.1140/epjp/s13360-021-02152-x} {\bibfield  {journal}
  {\bibinfo  {journal} {The European Physical Journal Plus}\ }\textbf {\bibinfo
  {volume} {136}},\ \bibinfo {pages} {1157} (\bibinfo {year}
  {2021})}\BibitemShut {NoStop}%
\bibitem [{\citenamefont {Jos\'e}\ \emph {et~al.}(1977)\citenamefont {Jos\'e},
  \citenamefont {Kadanoff}, \citenamefont {Kirkpatrick},\ and\ \citenamefont
  {Nelson}}]{JKKN77}%
  \BibitemOpen
  \bibfield  {author} {\bibinfo {author} {\bibfnamefont {J.~V.}\ \bibnamefont
  {Jos\'e}}, \bibinfo {author} {\bibfnamefont {L.~P.}\ \bibnamefont
  {Kadanoff}}, \bibinfo {author} {\bibfnamefont {S.}~\bibnamefont
  {Kirkpatrick}},\ and\ \bibinfo {author} {\bibfnamefont {D.~R.}\ \bibnamefont
  {Nelson}},\ }\bibfield  {title} {\bibinfo {title} {Renormalization, vortices,
  and symmetry-breaking perturbations in the two-dimensional planar model},\
  }\href {https://doi.org/10.1103/PhysRevB.16.1217} {\bibfield  {journal}
  {\bibinfo  {journal} {Phys. Rev. B}\ }\textbf {\bibinfo {volume} {16}},\
  \bibinfo {pages} {1217} (\bibinfo {year} {1977})}\BibitemShut {NoStop}%
\bibitem [{\citenamefont {Manousakis}(1991)}]{Manousakis91}%
  \BibitemOpen
  \bibfield  {author} {\bibinfo {author} {\bibfnamefont {E.}~\bibnamefont
  {Manousakis}},\ }\bibfield  {title} {\bibinfo {title} {The
  spin-\textonehalf{} heisenberg antiferromagnet on a square lattice and its
  application to the cuprous oxides},\ }\href
  {https://doi.org/10.1103/RevModPhys.63.1} {\bibfield  {journal} {\bibinfo
  {journal} {Rev. Mod. Phys.}\ }\textbf {\bibinfo {volume} {63}},\ \bibinfo
  {pages} {1} (\bibinfo {year} {1991})}\BibitemShut {NoStop}%
\bibitem [{\citenamefont {Fr\"ohlich}\ and\ \citenamefont
  {Lieb}(1977)}]{Frohlich77}%
  \BibitemOpen
  \bibfield  {author} {\bibinfo {author} {\bibfnamefont {J.}~\bibnamefont
  {Fr\"ohlich}}\ and\ \bibinfo {author} {\bibfnamefont {E.~H.}\ \bibnamefont
  {Lieb}},\ }\bibfield  {title} {\bibinfo {title} {Existence of phase
  transitions for anisotropic heisenberg models},\ }\href
  {https://doi.org/10.1103/PhysRevLett.38.440} {\bibfield  {journal} {\bibinfo
  {journal} {Phys. Rev. Lett.}\ }\textbf {\bibinfo {volume} {38}},\ \bibinfo
  {pages} {440} (\bibinfo {year} {1977})}\BibitemShut {NoStop}%
\bibitem [{\citenamefont {Kubo}\ and\ \citenamefont {Kishi}(1988)}]{Kubo88}%
  \BibitemOpen
  \bibfield  {author} {\bibinfo {author} {\bibfnamefont {K.}~\bibnamefont
  {Kubo}}\ and\ \bibinfo {author} {\bibfnamefont {T.}~\bibnamefont {Kishi}},\
  }\bibfield  {title} {\bibinfo {title} {Existence of long-range order in the
  $\mathrm{XXZ}$ model},\ }\href {https://doi.org/10.1103/PhysRevLett.61.2585}
  {\bibfield  {journal} {\bibinfo  {journal} {Phys. Rev. Lett.}\ }\textbf
  {\bibinfo {volume} {61}},\ \bibinfo {pages} {2585} (\bibinfo {year}
  {1988})}\BibitemShut {NoStop}%
\bibitem [{\citenamefont {Ozeki}\ \emph {et~al.}(1989)\citenamefont {Ozeki},
  \citenamefont {Nishimori},\ and\ \citenamefont {Tomita}}]{Ozeki89}%
  \BibitemOpen
  \bibfield  {author} {\bibinfo {author} {\bibfnamefont {Y.}~\bibnamefont
  {Ozeki}}, \bibinfo {author} {\bibfnamefont {H.}~\bibnamefont {Nishimori}},\
  and\ \bibinfo {author} {\bibfnamefont {Y.}~\bibnamefont {Tomita}},\
  }\bibfield  {title} {\bibinfo {title} {Long-range order in antiferromagnetic
  quantum spin systems},\ }\href {https://doi.org/10.1143/JPSJ.58.82}
  {\bibfield  {journal} {\bibinfo  {journal} {Journal of the Physical Society
  of Japan}\ }\textbf {\bibinfo {volume} {58}},\ \bibinfo {pages} {82}
  (\bibinfo {year} {1989})},\ \Eprint
  {https://arxiv.org/abs/https://doi.org/10.1143/JPSJ.58.82}
  {https://doi.org/10.1143/JPSJ.58.82} \BibitemShut {NoStop}%
\bibitem [{\citenamefont {Barnes}\ \emph {et~al.}(1989)\citenamefont {Barnes},
  \citenamefont {Kotchan},\ and\ \citenamefont {Swanson}}]{Barnes89}%
  \BibitemOpen
  \bibfield  {author} {\bibinfo {author} {\bibfnamefont {T.}~\bibnamefont
  {Barnes}}, \bibinfo {author} {\bibfnamefont {D.}~\bibnamefont {Kotchan}},\
  and\ \bibinfo {author} {\bibfnamefont {E.~S.}\ \bibnamefont {Swanson}},\
  }\bibfield  {title} {\bibinfo {title} {Evidence for a phase transition in the
  zero-temperature anisotropic two-dimensional heisenberg antiferromagnet},\
  }\href {https://doi.org/10.1103/PhysRevB.39.4357} {\bibfield  {journal}
  {\bibinfo  {journal} {Phys. Rev. B}\ }\textbf {\bibinfo {volume} {39}},\
  \bibinfo {pages} {4357} (\bibinfo {year} {1989})}\BibitemShut {NoStop}%
\bibitem [{\citenamefont {Ding}(1990)}]{Ding90b}%
  \BibitemOpen
  \bibfield  {author} {\bibinfo {author} {\bibfnamefont {H.~Q.}\ \bibnamefont
  {Ding}},\ }\bibfield  {title} {\bibinfo {title} {Antiferromagnetic
  transitions in high-tc materials},\ }\href
  {https://doi.org/10.1088/0953-8984/2/39/010} {\bibfield  {journal} {\bibinfo
  {journal} {Journal of Physics: Condensed Matter}\ }\textbf {\bibinfo {volume}
  {2}},\ \bibinfo {pages} {7979} (\bibinfo {year} {1990})}\BibitemShut
  {NoStop}%
\bibitem [{\citenamefont {Ding}(1992{\natexlab{b}})}]{Ding92b}%
  \BibitemOpen
  \bibfield  {author} {\bibinfo {author} {\bibfnamefont {H.-Q.}\ \bibnamefont
  {Ding}},\ }\bibfield  {title} {\bibinfo {title} {Could in-plane exchange
  anisotropy induce the observed antiferromagnetic transitions in the undoped
  high-tc materials?},\ }\href {https://doi.org/10.1103/physrevlett.68.1927}
  {\bibfield  {journal} {\bibinfo  {journal} {Physical Review Letters}\
  }\textbf {\bibinfo {volume} {68}},\ \bibinfo {pages} {1927} (\bibinfo {year}
  {1992}{\natexlab{b}})}\BibitemShut {NoStop}%
\bibitem [{\citenamefont {Aplesnin}(1998)}]{Aplesnin98}%
  \BibitemOpen
  \bibfield  {author} {\bibinfo {author} {\bibfnamefont {S.~S.}\ \bibnamefont
  {Aplesnin}},\ }\bibfield  {title} {\bibinfo {title} {Quantum monte carlo
  analysis of the 2d heisenberg antiferromagnet with s = 1/2: the influence of
  exchange anisotropy},\ }\href {https://doi.org/10.1088/0953-8984/10/44/012}
  {\bibfield  {journal} {\bibinfo  {journal} {Journal of Physics: Condensed
  Matter}\ }\textbf {\bibinfo {volume} {10}},\ \bibinfo {pages} {10061}
  (\bibinfo {year} {1998})}\BibitemShut {NoStop}%
\bibitem [{\citenamefont {Lin}\ \emph {et~al.}(2001)\citenamefont {Lin},
  \citenamefont {Flynn},\ and\ \citenamefont {Betts}}]{lin01}%
  \BibitemOpen
  \bibfield  {author} {\bibinfo {author} {\bibfnamefont {H.-Q.}\ \bibnamefont
  {Lin}}, \bibinfo {author} {\bibfnamefont {J.~S.}\ \bibnamefont {Flynn}},\
  and\ \bibinfo {author} {\bibfnamefont {D.~D.}\ \bibnamefont {Betts}},\
  }\bibfield  {title} {\bibinfo {title} {Exact diagonalization and quantum
  monte carlo study of the spin-$\frac{1}{2}$ $\mathrm{XXZ}$ model on the
  square lattice},\ }\href {https://doi.org/10.1103/PhysRevB.64.214411}
  {\bibfield  {journal} {\bibinfo  {journal} {Phys. Rev. B}\ }\textbf {\bibinfo
  {volume} {64}},\ \bibinfo {pages} {214411} (\bibinfo {year}
  {2001})}\BibitemShut {NoStop}%
\bibitem [{\citenamefont {Roscilde}\ \emph {et~al.}(2003)\citenamefont
  {Roscilde}, \citenamefont {Cuccoli},\ and\ \citenamefont
  {Verrucchi}}]{Roscilde03}%
  \BibitemOpen
  \bibfield  {author} {\bibinfo {author} {\bibfnamefont {T.}~\bibnamefont
  {Roscilde}}, \bibinfo {author} {\bibfnamefont {A.}~\bibnamefont {Cuccoli}},\
  and\ \bibinfo {author} {\bibfnamefont {P.}~\bibnamefont {Verrucchi}},\
  }\bibfield  {title} {\bibinfo {title} {Phase transitions in anisotropic
  two-dimensional quantum antiferromagnets},\ }\href
  {https://doi.org/https://doi.org/10.1002/pssb.200301697} {\bibfield
  {journal} {\bibinfo  {journal} {physica status solidi (b)}\ }\textbf
  {\bibinfo {volume} {236}},\ \bibinfo {pages} {433} (\bibinfo {year}
  {2003})},\ \Eprint
  {https://arxiv.org/abs/https://onlinelibrary.wiley.com/doi/pdf/10.1002/pssb.200301697}
  {https://onlinelibrary.wiley.com/doi/pdf/10.1002/pssb.200301697} \BibitemShut
  {NoStop}%
\bibitem [{\citenamefont {Cuccoli}\ \emph {et~al.}(2003)\citenamefont
  {Cuccoli}, \citenamefont {Roscilde}, \citenamefont {Tognetti}, \citenamefont
  {Vaia},\ and\ \citenamefont {Verrucchi}}]{Cuccoli03}%
  \BibitemOpen
  \bibfield  {author} {\bibinfo {author} {\bibfnamefont {A.}~\bibnamefont
  {Cuccoli}}, \bibinfo {author} {\bibfnamefont {T.}~\bibnamefont {Roscilde}},
  \bibinfo {author} {\bibfnamefont {V.}~\bibnamefont {Tognetti}}, \bibinfo
  {author} {\bibfnamefont {R.}~\bibnamefont {Vaia}},\ and\ \bibinfo {author}
  {\bibfnamefont {P.}~\bibnamefont {Verrucchi}},\ }\bibfield  {title} {\bibinfo
  {title} {Quantum monte carlo study of $s=\frac{1}{2}$ weakly anisotropic
  antiferromagnets on the square lattice},\ }\href
  {https://doi.org/10.1103/PhysRevB.67.104414} {\bibfield  {journal} {\bibinfo
  {journal} {Phys. Rev. B}\ }\textbf {\bibinfo {volume} {67}},\ \bibinfo
  {pages} {104414} (\bibinfo {year} {2003})}\BibitemShut {NoStop}%
\end{thebibliography}%

\end{document}